\documentclass{aa}  

\usepackage{graphicx}
\usepackage{txfonts}
\usepackage{hyperref}
\usepackage{xcolor}
\usepackage{subcaption}
\usepackage{float}
\usepackage{soul}

\def\Msun{\rm M_\odot}
\def\msun{{\Msun}}

\def\Msun{\rm M_\odot}

\def\HI{\hbox{H~$\scriptstyle\rm I\ $}} 

\begin{document}

   \title{Astraeus X: Indications of a top-heavy initial mass function in highly star-forming galaxies from JWST observations at $z>10$}
   \titlerunning{Top-heavy IMF in highly star-forming galaxies}

   \author{Anne~Hutter\inst{1,2}\fnmsep\thanks{\email{anne.hutter@nbi.ku.dk}}, 
            Elie~R.~Cueto\inst{1},
            Pratika~Dayal\inst{3},
            Stefan~Gottl\"ober\inst{4},
            Maxime~Trebitsch\inst{5},
            Gustavo~Yepes\inst{6,7}
          }
    \authorrunning{Hutter et al.}

   \institute{Niels Bohr Institute, University of Copenhagen, Jagtvej 128, DK-2200, Copenhagen N, Denmark
         \and Cosmic Dawn Center (DAWN)
         \and Kapteyn Astronomical Institute, University of Groningen, P.O. Box 800, 9700 AV Groningen, The Netherlands
         \and Leibniz-Institut f\"ur Astrophysik, An der Sternwarte 16, 14482 Potsdam, Germany
         \and Sorbonne Universit\'e, Observatoire de Paris, PSL Research University, CNRS, LERMA, 75014 Paris, France
         \and Departamento de Fisica Teorica, Modulo 8, Facultad de Ciencias, Universidad Autonoma de Madrid, 28049 Madrid, Spain
         \and CIAFF, Facultad de Ciencias, Universidad Autonoma de Madrid, 28049 Madrid, Spain
         }

   \date{Received - -, -; accepted - -, -}

 
  \abstract
   {The James Webb Space Telescope (JWST) has uncovered an abundance of $z>10$ galaxies bright in the ultraviolet (UV) that has challenged traditional theoretical models at high redshifts. Various new models have recently emerged to address this discrepancy by refining their description of star formation.}
   {Here, we investigate whether modifications to the stellar initial mass function (IMF) alone can reproduce the $z>10$ UV luminosity functions (UV LFs) when the star formation rate is used as a proxy for the fraction of massive stars.}
   {We incorporate an Evolving IMF into the {\sc astraeus} galaxy evolution and reionisation simulation framework, which becomes increasingly top-heavy as the gas density in a galaxy rises above a given threshold. Our implementation accounts for the IMF's effects on supernova (SN) feedback, metal enrichment, and UV and ionising emissivities.}
   {For this Evolving IMF model, we find that 
   (i) the maximum UV luminosity enhancement is twice as large in massive galaxies ($\Delta M_\mathrm{UV}\simeq2.6$) than those where star formation is strongly limited by SN feedback ($\Delta M_\mathrm{UV}\simeq1.3$);
   (ii) it successfully reproduces the observed UV LFs at $z=5-15$;
   (iii) galaxies with top-heavy IMFs exhibit the highest star formation rates, driven by their location in local density peaks, which facilitates higher gas accretion rates;
   (iv) the $1\sigma$ variances in the UV luminosity are only slightly higher compared to when assuming a Salpeter IMF, but the $2\sigma$ variances are significantly increased by a factor of $1.4-2$ boosting the abundance of UV-bright galaxies at $z>10$;
   (v) reionisation begins earlier with more extended large ionised regions and fewer smaller ones during its initial stages, though these differences diminish at lower redshifts, leading to a similar end of reionisation at $z\simeq5.6$.}
   {}


   \maketitle
%
\section{Introduction}
\label{sec_introduction}

The first galaxies emerging in our Universe during its first billion years ushered the last major phase transition, the Epoch of Reionisation (EoR), by emitting energetic photons that gradually ionised the neutral hydrogen in the intergalactic medium (IGM) \citep{Barkana2001, Dayal2018}. Exactly how this phase transition unfolded on global and local scales at $z\simeq20-5$ remains unconstrained, given our incomplete understanding of the population and properties of the first galaxies. Detections of numerous $z\gtrsim8$ galaxies with the James Webb Space Telescope (JWST) have revealed young, star-forming, compact, metal-poor galaxies whose stellar populations emit harder hydrogen-ionising radiation during the EoR \citep[e.g.][]{Adams2023, Atek2024, Bradley2023, Bunker2023, CurtisLake2023, Heintz2023a, Rinaldi2024, Schaerer2022, Simmonds2024a, Vanzella2023}. However, these JWST observations also present a puzzle: the detection of a higher abundance of UV-bright galaxies at $z>10$ than standard theoretical models predict \citep[e.g.][]{Labbe2023, Adams2023, Adams2024, ArrabalHaro2023, Atek2023, Austin2023, BoylanKolchin2023, Donnan2024, Harikane2024, McLeod2024}. This puzzle remains even when limiting the galaxy candidates to those spectroscopically confirmed \citep[see][]{Harikane2024}.

Various theoretical investigations have been conducted to explain this puzzle, focusing on selection biases and altered physical processes. Some studies suggest that the observed galaxies are biased samples due to the stochastic nature of star formation, mostly detecting galaxies undergoing periods of intense star formation while missing those with strongly suppressed or no star formation \citep[e.g.][]{Mason2023, Mirocha2023, Shen2023, Sun2023, Kravtsov2024, Gelli2024}. However, assuming a dark matter (DM) mass-dependent UV luminosity variance, consistent with the {\sc fire} simulation and scaling inversely with the halo's escape velocity ($\propto M_h^{-1/3}$), has been shown to be insufficient in reproducing the abundance of UV-bright galaxies at $z>10$ \citep{Gelli2024}. This could indicate that the physical conditions and processes at very high redshifts differ. Indeed, the compactness of these early galaxies implies higher densities of metal-poor gas, where stellar feedback might be less efficient in suppressing subsequent star formation, thereby boosting the star formation efficiency (SFE) \citep{Dekel2023}. Additionally, radiatively driven outflows may eject dust from star formation sites, reducing UV attenuation by dust \citep{Ferrara2023, Fiore2023, Mauerhofer2023, Yung2023, Ziparo2023}, or massive black holes accreting at or slightly above the Eddington rate could contribute to the observed UV luminosity \citep{Pacucci2022}. However, testing whether the SFE is enhanced in dense, metal-poor star-forming clouds requires radiative hydrodynamical simulations, and even state-of-the-art semi-analytic galaxy evolution models fail to reproduce the abundance of $z>10$ UV-bright galaxies when neglecting dust attenuation \citep[see e.g.][]{Cueto2024}.

Theoretical models and simulations indicate that the first metal-free stars were massive, with stellar masses of $\gtrsim60\msun$ (\citealt{Abel2002, Bromm2002, Yoshida2006, Fukushima2020}, although see \citealt{Clark2011}), suggesting that the stellar initial mass function (IMF) in early galaxies was skewed towards higher masses, thereby reducing the overall mass-to-light ratio. Several observations seem to support this idea, with high-redshift galaxies exhibiting higher ionising emissivities \citep{Simmonds2024a}, bluer UV slopes \citep{Cullen2024}, and unusual abundance ratios -- such as enhanced [N/O] \citep[e.g. in GN-z11, CEERS-1019, GLASS-150008, JADES-GS-z9;][]{Bunker2023, Cameron2024, Isobe2023, Curti2024, Senchyna2024, Topping2024, Watanabe2024}, lower [C/O] \citep[e.g. in CEERS-1019;][]{Isobe2023}, and low [O/Fe] abundances \citep[e.g. in GN-z11;][]{Nakane2024}. 
Enhanced [N/O] abundances, which indicate active CNO cycling in stars \citep{Ekstrom2021}, are expected in supermassive stars ($\gtrsim1000\msun$) \citep{Charbonnel2023, Nagele2023, Isobe2023}, very massive stars (100-1000$\msun$) \citep{Vink2023, Curti2024}, rotating massive stars \citep{Nandal2024}, Wolf-Rayet stars in scenarios involving dual starbursts \citep{Kobayashi2024} or tidal disruptions. Similarly, lower [C/O] abundances are consistent with contributions from supermassive stars, Wolf-Rayet stars, or tidal disruptions \citep{Isobe2023, Watanabe2024a}. 
These observational findings suggest that the stellar populations in the first galaxies formed with an IMF that was more top-heavy than the local Salpeter IMF \citep{Salpeter}. First estimates of the associated reduction in the stellar mass-to-UV luminosity ratio have concluded that a top-heavy IMF could resolve the UV-bright galaxy abundance problem at $z>10$ \citep{Haslbauer2022, Trinca2024, Yung2023}.

However, these works omit the IMF dependency of stellar feedback, neglecting the enhanced fraction of stars exploding as supernovae (SNe), which reduces not only subsequent star formation more immediately but also increases and alters the metal enrichment of the interstellar gas per stellar mass formed. In \citet{Cueto2024}, we addressed this gap using the {\sc astraeus} framework, a semi-numerical model that tracks the interdependent evolution of galaxies and reionisation \citep{Hutter2021a, Ucci2023, Hutter2023a}. We adopted the IMF inferred from star-forming cloud simulations in \citet{Chon2022}, where the IMF became increasingly top-heavy with rising redshift and decreasing gas-phase metallicity - a result of the increasing temperature of the Cosmic Microwave Background at higher redshifts and the associated decreasing gas cooling efficiency.
To reproduce the observed UV luminosity functions (UV LFs) at $z\lesssim10$, we adjusted the free model parameters in {\sc astraeus} to adopt an overall lower SFE than for the standard assumption of a constant Salpeter IMF. This reduction in SFE roughly corresponded to the decrease caused by enhanced radiative pressure from young stars, as reported by \citet{Menon2024}, for cloud surface densities of $\sim10^3~\msun$~pc$^{-1}$ at $Z=0.01$~Z$_\odot$ and a UV luminosity enhancement of $10$, and compensated for the fact that {\sc astraeus} did not capture the effects of radiative feedback from massive stars on star formation. Yet, the resulting UV LFs at $z=5-15$ remained consistent with those produced by models assuming a constant Salpeter IMF. 
In contrast, \citet{Lu2025} does not alter the SFE and finds that adopting the same top-heavy IMF for all galaxies can match the observed UV LF at $z\lesssim12$.

While \citet{Cueto2024} assumed top-heavy IMFs for all galaxies, and \citet{Lu2025} applied a universal top-heavy IMF to galaxies in the burst mode of star formation within the {\sc galform} model, we consider here a scenario where only a fraction of galaxies have a top-heavy IMF, with its degree of top-heaviness varying according to each galaxy's properties. Observations of local star-forming regions and nearby galaxies find the IMF to become more top-heavy in regions of intense star formation and dense gas environments \citep[e.g.][]{Meurer2009, Watts2018, Zhang2018, Gunawardhana2011, Fontanot2017, Fontanot2018a}, a trend possibly also due to the higher cosmic ray densities in these regions raising gas temperatures \citep{Papadopoulos2011}. For this reason and because the radiation and elemental abundances in $z>5$ galaxies suggest a higher prevalence of massive stars in UV-bright galaxies, we investigate a scenario where the IMF becomes increasingly top-heavy with rising specific star formation rate (sSFR), which we refer to as the Evolving IMF. Naturally, this IMF dependency also aligns with those found in the galaxy-wide IMF (gwIMF) theory \citep{Kroupa2003, Weidner2013, Yan2017, Jerabkova2018} that relies on empirical relations, such as the relation between the maximum star mass and cluster mass. 
Assuming that the SFE is unaffected by radiative feedback from massive stars, as suggested for higher cloud densities \citep{Menon2024}, we primarily investigate whether adapting the otherwise assumed Salpeter IMF to a more top-heavy IMF in galaxies with higher sSFR values can reproduce the observed UV LFs at $z=5-15$ without altering the SFE or SNe wind coupling efficiencies. To further understand the results and their implications, we also explore (i) the extent to which a top-heavy IMF enhances the UV luminosity of galaxies, regardless of whether star formation is suppressed by SN feedback; (ii) the properties and characteristics of galaxies with a top-heavy IMF; and (iii) how reionisation changes when transitioning from a universal Salpeter IMF to our Evolving IMF.

This paper is organised as follows. In Section \ref{sec_model}, we describe the updates made to the {\sc astraeus} model compared to \citet{Cueto2024}, focusing on our new parameterisation of the IMF and its scaling with galactic properties to mimic a dependency on the sSFR. Section \ref{sec_UVboost} introduces a simple toy model to explore the maximum possible UV luminosity boost for various halo masses when assuming our top-heavy IMF parameterisation. We then discuss in Section \ref{sec_uvlfs} how the free model parameters in our Evolving IMF parameterisation affect the UV LFs and identify the parameter set that best fits the UV LFs at $z=5-15$. In Section \ref{sec_galpop} and \ref{sec_UVlum_scatter}, we analyse the characteristics of the galaxies with top-heavy IMFs in our model and examine changes in the UV luminosity - halo mass distribution. We briefly discuss the impact of the Evolving IMF on reionisation in Section \ref{sec_reionisation} and the range of the ionising emissivity values of high-redshift galaxies in Section \ref{sec_xion}. We conclude in Section \ref{sec_conclusions}. Throughout this paper, we assume the AB magnitude system \citep{Oke1983} and a $\Lambda$CDM universe with the \citet{planck2016} cosmological parameters:  $\Omega_\Lambda=0.692885$, $\Omega_m=0.307115$, $\Omega_b=0.048206$, $H_0=100h=67.77$km~s$^{-1}$Mpc$^{-1}$, $n_s=0.96$, and $\sigma_8=0.8228$.

\section{The model}
\label{sec_model}

In this Section, we briefly describe the changes we have incorporated in the {\sc astraeus} framework to follow an IMF that becomes increasingly top-heavy as the gas content and sSFR in galaxies rise. The {\sc astraeus} framework post-processes the {\sc vsmdpl} (very small multidark planck) DM-only N-body simulation (for details, see Tab.~\ref{tab_vsmdpl}) with an enhanced version of the semi-analytic galaxy evolution model {\sc delphi} \citep{Dayal2014, Dayal2022} that is coupled to the semi-numerical reionisation scheme {\sc cifog} \citep{Hutter2018a} in a fully self-consistent manner. {\sc astraeus} incorporates all the key physical processes that shape the evolution of $z>5$ galaxies, including gas accretion, mergers, star formation, SN feedback, metal enrichment, dust formation, and radiative feedback from reionisation. It also models reionisation, tracking the time evolution of the spatial distribution of ionised regions. A detailed description of the implementation of these processes can be found in \citet{Hutter2021a, Hutter2018a, Ucci2023, Hutter2023a} and \citet{Cueto2024}. 
We advance the version described in \citet{Cueto2024} by adjusting the parameterisation of the IMF and the ionising and ultraviolet luminosities as we detail in this Section. We note that to avoid artificial starbursts of galaxies coming into existence in the simulation (i.e. galaxies without progenitors) - particularly since our new IMF parameterisation will depend on the gas-to-DM mass ratio - we change the assumption on the initial gas mass in a galaxy upon its existence (i.e. when it has no progenitors). Instead of assuming $M_\mathrm{g}^i=f_g\frac{\Omega_b}{\Omega_m}M_h$, we adopt now $M_\mathrm{g}^i=0~\msun$. Since galaxies without progenitors reside in DM halos with $M_h\lesssim10^{8.3}\msun$, their SN feedback limited star formation is highly stochastic \citep{Legrand2022} oscillating between phases of forming stars and forming no stars; we assume these galaxies to start in a phase of no star formation.\footnote{The exact value adopted for $M_\mathrm{g}^i$ has no significant effect on galaxies with converged star formation histories (SFHs, $M_h\gtrsim10^{8.6}\msun$, see Appendix B in \citealt{Hutter2021a}).}

\begin{table}
    \centering
    \caption{Details of the DM-only N-body simulation {\sc vsmdpl}}
    \label{tab_vsmdpl}
    \begin{tabular*}{\columnwidth}{ccc}
        \hline
        Parameter & Value & Description \\
        \hline
        \hline
        $L$ & $160h^{-1}$~cMpc & simulation box size\\
        $M_\mathrm{h, min}$ & $10^{8.2}h^{-1}~\msun$ & minimum halo mass\\
        \hline
    \end{tabular*}
\end{table}

\subsection{The initial mass function}
\label{subsec_imf}

\begin{figure}
    \centering
    \includegraphics[width=\hsize]{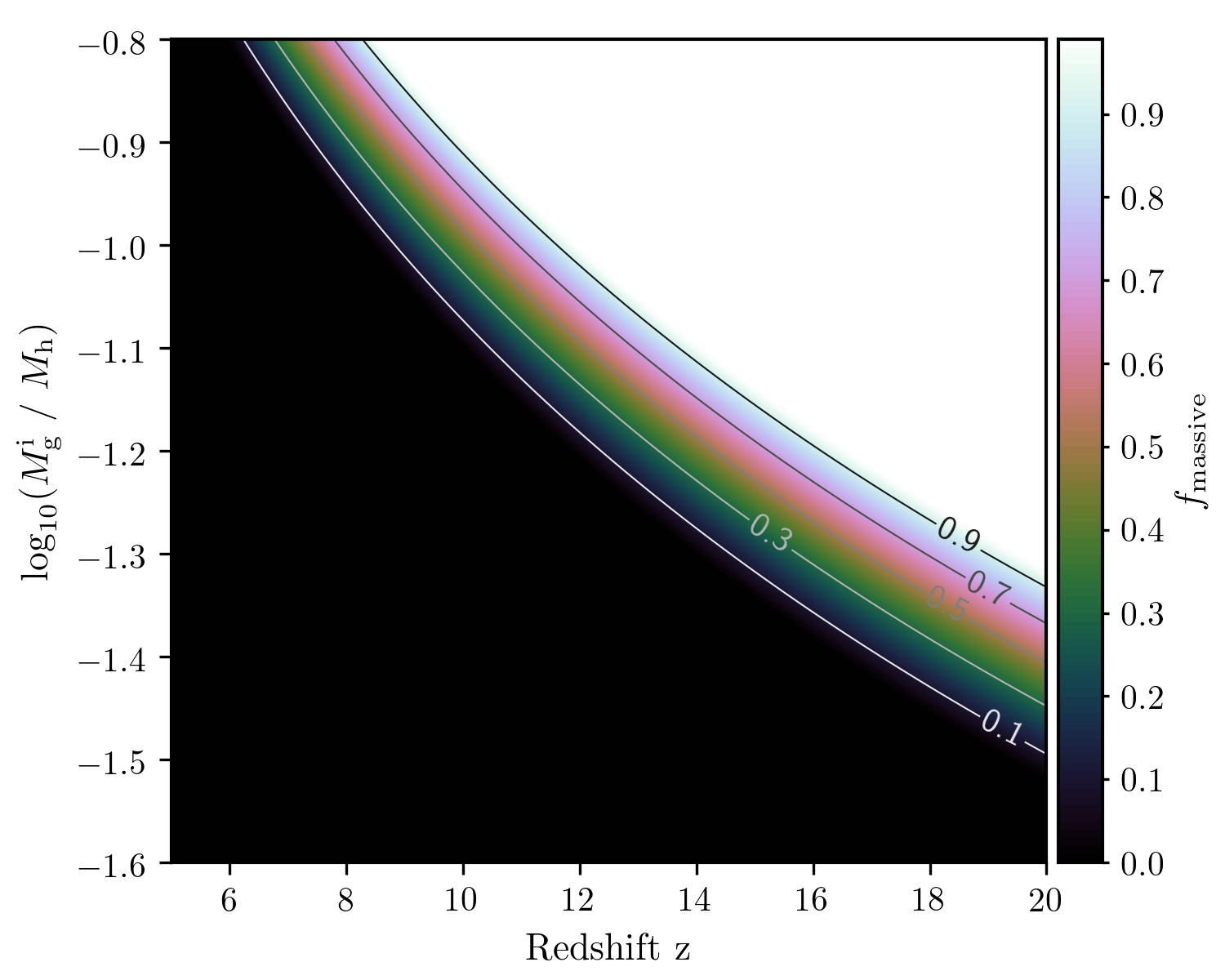}
    \caption{The dependence of $f_\mathrm{massive}$ on redshift $z$ and the gas-to-DM mass ratio for the Evolving IMF model. The cosmological baryon ratio $\Omega_b/\Omega_m$ corresponds to $\sim-0.8$.}
    \label{fig_IMF}
\end{figure}

We adopt an IMF with a Salpeter slope of $\gamma=-2.35$ for stars with masses below a mass $M_\mathrm{c}$ and a log-flat slope of $-1$ for stars with masses above $M_\mathrm{c}$, consistent with the findings in \citet{Chon2022}.
\begin{eqnarray}
    \frac{\mathrm{d}N}{\mathrm{d}M} = 
    \begin{cases}
        \left(1 - f_\mathrm{massive}\right)\ \frac{\gamma + 2}{M_\mathrm{c}^{\gamma +2}-M_\mathrm{i}^{\gamma+2}}\ M^{\gamma} & ,\ M_\mathrm{i} \leq M < M_\mathrm{c}\\
        f_\mathrm{massive}\ \frac{1}{M_\mathrm{f} - M_\mathrm{c}}\ M^{-1} & ,\ M_\mathrm{c} \leq M \leq M_\mathrm{f}.
    \end{cases}
    \label{eq_evolvingIMF}
\end{eqnarray}
$M_\mathrm{c}$ represents the stellar mass at which the IMF slope changes, and its value depends on the fraction of mass formed in massive stars, $f_\mathrm{massive}$. We numerically derive the relation between $f_\mathrm{massive}$ and $M_\mathrm{c}$ by solving the equation $\int_{M_\mathrm{i}}^{M_\mathrm{f}} \frac{\mathrm{d}N}{\mathrm{d}M}~ M~ \mathrm{d}M = 1 \msun$.
However, in contrast to \citet{Cueto2024}, we scale $f_\mathrm{massive}$ with the gas content of a galaxy above a redshift-dependent gas density threshold. This scaling and threshold are chosen such that $f_\mathrm{massive}$ is effectively proportional to the sSFR of a galaxy above a specified threshold, $\mathrm{sSFR}_\mathrm{thresh}$.  
\begin{eqnarray}
    f_\mathrm{massive} &=& k\ \times\ \frac{\mathrm{sSFR} - \mathrm{sSFR}_\mathrm{thresh}}{\mathrm{sSFR}_\mathrm{max} - \mathrm{sSFR}_\mathrm{thresh}}
\end{eqnarray}
In our model, the SFE depends on the galaxy's gravitational potential. While in more massive galaxies, the SFE remains a constant value scaling with the gas free-fall time in halos, $\tau = (4\pi G \rho)^{-1/2} \propto H(z)^{-1}\propto(1+z)^{-3/2}$, at the respective redshift $z$, it decreases to an efficiency in lower mass galaxies where galaxies form only as many stars as are necessary to eject all their gas due to SN explosions,
\begin{eqnarray}
    f_\star^\mathrm{ej}(z) &=& \frac{v_c^2}{v_c^2 + f_w^\mathrm{eff}(z)\ E_{51} \nu_z} \left[  1 - \frac{f_w^\mathrm{eff}(z)\ E_{51} \sum_j \nu_j M_{\star,j}^\mathrm{new}(z_j)}{M_\mathrm{g}^i(z)~ v_c^2} \right].
    \label{eq_fstarEj}
\end{eqnarray}
Here, $v_c$ represents the halo's rotational velocity, $E_{51}$ denotes the energy released by a Type II supernova (SNII), and $\nu_z$ is the IMF-dependent fraction of stellar mass that forms and explodes within the current time step. For the stellar mass formed in the previous time step $j$, $M_{\star,j}^\mathrm{new}$, $\nu_j$ represents the IMF-dependent fraction that explodes in the current time step.
\begin{eqnarray}
    f_w^\mathrm{eff}(z) &=& \frac{f_w}{1 + \left( \frac{\Delta t}{20~\mathrm{Myr}} - 1\right) \frac{M_g^\mathrm{mer}(z)}{M_g^i(z)}}
    \label{eq_fweff}
\end{eqnarray}
describes the fraction of SN energy coupling to the gas and driving winds. $M_\star^\mathrm{new}$ is the stellar mass formed in the current time step with length $\Delta t$, while $M_\mathrm{g}^i$ denotes the gas mass at the beginning of a time step and $M_\mathrm{g}^\mathrm{mer}$ the fraction of gas inherited by the galaxy's progenitors. The SFE is thus given by
\begin{eqnarray}
    f_\star^\mathrm{eff}(z) = \min\left[f_\star^\mathrm{ej}(z), f_\star \frac{\tau(z=9)}{\tau(z)} \frac{\Delta t}{20~\mathrm{Myr}}\right].
    \label{eq_feff}
\end{eqnarray}
We note that, although the choice of the normalisation redshift for $f_\star$ is arbitrary, we choose $z=9$ as it corresponds to the highest redshifts where standard IMFs reproduce the observed UV LF. From the above, it is evident that the SFE and thus the sSFR depend on the assumed IMF. For this reason, we focus on the factors that primarily influence the sSFRs in galaxies with efficient star formation, namely the available gas mass for star formation, $M_\mathrm{g}^i$, and the redshift $z$.\footnote{We note that this approximation becomes inaccurate as star formation is increasingly SN feedback-limited.}
\begin{eqnarray}
    \mathrm{sSFR} &=& \frac{\mathrm{SFR}}{M_\star} = \frac{f_\star^\mathrm{eff} M_\mathrm{g}^i}{\Delta t ~ M_\star} \propto \frac{M_\mathrm{g}^i}{M_h} (1+z)^{3/2}, 
    \label{eq_sSFR}
\end{eqnarray}
Here SFR is the star formation rate and $M_\star$ and $M_h$ the stellar and halo masses of the galaxy at $z$.
Based on this proportionality, we express the fraction of mass that is bound in massive stars as
\begin{eqnarray}
    f_\mathrm{massive} &=& \min \left[ \frac{k x}{(1+z_\mathrm{max})^{3/2} - (1+z_\mathrm{thresh})^{3/2}}\times~ H \left(x \right), 1 \right] \nonumber\\
    x &=& \frac{M_\mathrm{g}^i \Omega_\mathrm{m}}{M_\mathrm{h} \Omega_\mathrm{b}} (1+z)^{3/2} - (1+z_\mathrm{thresh})^{3/2} 
    \label{eq_fmassive}
\end{eqnarray}
with $k$, $z_\mathrm{thresh}$ and $z_\mathrm{max}$ being free model parameters and $H(x)$ the Heaviside function. Eqn.~\ref{eq_fmassive} shows that $f_\mathrm{massive}$ increases towards higher redshifts and higher gas-to-DM mass ratios, resulting in an increasingly top-heavy IMF. However, if $k$ is large, $f_\mathrm{massive}$ stalls at $1$ for sufficiently high gas-to-DM ratios, corresponding to a log-flat IMF. For $z_\mathrm{max}=20$, $z_\mathrm{thresh}=6$, and $k=7$, the threshold specific star formation rate, $\mathrm{sSFR}\mathrm{thresh}$, is approximately $25$~Gyr$^{-1}$, and $f\mathrm{massive}$ approaches $1$ as the sSFR increases to around $60$~Gyr$^{-1}$. In Fig.~\ref{fig_IMF} we show how the corresponding $f_\mathrm{massive}$ values depends on redshift and the gas-to-DM mass ratio of each galaxy.
We note that in the remainder of this paper, we refer to an IMF that is top-heavier than the Salpeter IMF as top-heavy.

\subsection{The ionising and ultraviolet luminosities}
\label{subsec_luminosities}

As described in \citet{Cueto2024}, we use the stellar population synthesis code {\sc starburst99} \citep{Leitherer1999} to generate the spectra for the various IMFs with $f_\mathrm{massive}$ values ($[0,1]$), metallicities ($Z=[0.001, 0.008]$) and ages ($t=[10^{5}~\mathrm{yr},~10^{9}~\mathrm{yr}]$). From the simulated spectra, we derive and fit the time evolution of the ionising (number of photons with energies $>13.6$~eV per second) and UV (at $1500$\AA) luminosities for each $(f_\mathrm{massive}, Z)$ combination. The analytical fitting functions for $N_\mathrm{ion}$ $[\mathrm{s}^{-1}]$ and $L_\mathrm{UV}$ $[\mathrm{erg~s}^{-1}\mathrm{\AA}^{-1}]$ are listed in Appendix \ref{app_fits}. We note that these fits differ from those presented in \citet{Cueto2024}, as we use $f_\mathrm{massive}$ and $Z$ as scaling parameters instead of $z$ and $Z$, and we adopt a more complex fitting function for the ultraviolet luminosity (see Appendix \ref{app_fits}).

By convolving these fitting functions for the time evolution of the ionising and UV luminosities with each galaxy's star formation history (assuming continuous star formation within each timestep), we derive its intrinsic ionising emissivity, $\dot{Q}$, and intrinsic UV luminosity, $L_\mathrm{UV}^\mathrm{int}$ \citep[for details see][]{Hutter2021a, Hutter2023a}. 
We then model dust as graphite/carbonaceous grains and derive the UV dust attenuation using the galaxy's dust mass, $M_\mathrm{d}$, and spatial extension of the gas, $r_\mathrm{g}$. The optical depth is calculated as
\begin{eqnarray}
    \tau_\mathrm{UV,c} &=& \frac{3\sigma_d}{4as},
\end{eqnarray}
where $\sigma_d = \frac{M_d}{\pi r_d^2}$ is the dust surface density, $r_d$ is the dust distribution radius, and $a=0.03~\mu$m and $s=2.25$~g~cm$^{-3}$ are the radius and density of the graphite/carbonaceous dust grains, respectively. We assume gas and dust to be perfectly mixed, approximating the gas radius as $r_g=4.5\lambda r_\mathrm{vir}\left[ (1+z)/6\right]^{1.8}$ where $\lambda$ and $r_\mathrm{vir}$ represent the spin parameter and virial radius of the simulated DM halo. The final factor accounts for the redshift evolution of the compactness of galaxies, with gas occupying a larger fraction of halos at higher redshifts.\footnote{When this factor is omitted, $f_\star$ needs to be increased to $0.04$ to match the UV luminosity functions; while the UV LFs at $z\gtrsim10$ for the Salpeter IMF underpredict then the observations more significantly, the UV LFs for the Evolving IMF are hardly affected (see Sec.~\ref{subsec_bestfit_case} for explanation).} This assumption is motivated by ALMA observations showing that [C{\small II}] sizes for galaxies with fixed UV luminosity remain constant between $z\simeq7$ and $4$ \citep{Fujimoto2020, Fudamoto2022}. Using a slab-like geometry, we derive the UV continuum escape fraction as
\begin{eqnarray}
    f_\mathrm{esc}^\mathrm{c} &=& \frac{1 - e^{-\tau_\mathrm{UV,c}}}{\tau_\mathrm{UV,c}},
\end{eqnarray}
and calculate the observed UV luminosity of each galaxy as
\begin{eqnarray}
    L_\mathrm{UV} &=& f_\mathrm{esc}^\mathrm{c} L_\mathrm{UV}^\mathrm{int}.
\end{eqnarray}

\subsection{The different IMF scenarios}
\label{subsec_imf_models}

In the following, we consider two galaxy evolution models that only differ in their assumed initial mass function (IMF): a Salpeter IMF with a slope of $\gamma=-2.35$ covering stellar masses from $0.1~\msun$ to $100~\msun$, and the Evolving IMF described in Section \ref{subsec_imf}, which also covers stellar masses from $0.1~\msun$ to $100~\msun$. For both models, we assume the same star formation efficiencies, $f_\star$, and SN wind coupling efficiencies, $f_w$ (see Table \ref{tab_astraeus_param} for details).

\section{The UV luminosity boost for top-heavy IMFs}
\label{sec_UVboost}

\begin{figure}
    \centering
    \includegraphics[width=\hsize]{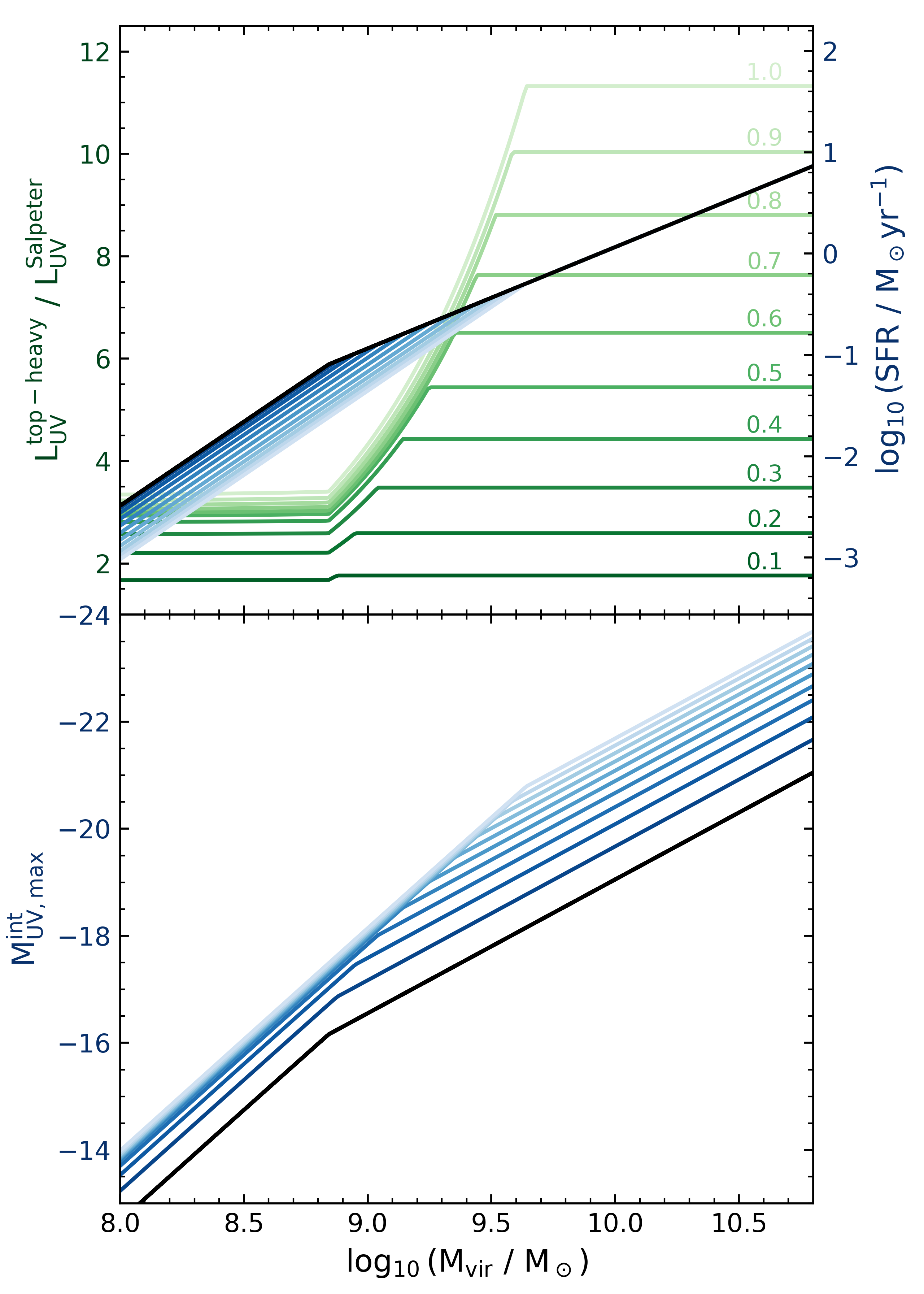}
    \caption{The halo mass dependence of the top-heavy IMF's increase in UV luminosity compared to the Salpeter IMF for a $30$~Myr long starburst (top panel, green lines), the SFR of gas-rich galaxies with top-heavy IMFs (top panel, blue lines) and their maximum UV luminosity (bottom panel, blue lines) for $f_\mathrm{massive}=0.1$, $0.2$, $0.3$, $0.4$, $0.5$, $0.6$, $0.7$, $0.8$, $0.9$, $1$ (from dark to light) at $z=13$. Black lines show the SFR and maximum intrinsic (dust unattenuated) UV luminosity of gas-rich galaxies with a Salpeter IMF. Assumed values are $f_\star'=0.027$ and $f_w=0.2$.}
    \label{fig_fmassive_dependence}
\end{figure}

Before analysing how the UV LFs and population of galaxies change for our Evolving IMF model, we consider a toy model for estimating the extent to which a more top-heavy IMF or increasing $f_\mathrm{massive}$ values can boost the UV luminosity of galaxies with a given halo mass.
For this purpose, we consider the SFE - fraction of gas mass forming stars over a given time interval $f_\star^\mathrm{eff}$ - when assuming instantaneous SN feedback in the {\sc astraeus} framework \citep[see][]{Hutter2021a}.  
Eqn.~\ref{eq_fstarEj} reduces then to
\begin{eqnarray}
    f_\star^\mathrm{ej} &=& \frac{v_c^2}{v_c^2 + f_w E_\mathrm{SN}}, 
\end{eqnarray}
with the rotational velocity of the halo being $v_c=\frac{G M_h}{r_h}(1+z) \propto M_h^{2/3} (1+z)$. We also define $f_\star' = f_\star \frac{\tau(z=9)}{\tau(z)} \frac{\Delta t}{20~\mathrm{Myr}}$. If we assume that the SN energy produced by stars forming with a top-heavy IMF ($E_\mathrm{SN}$) is related to the SN energy produced by stars forming with a Salpeter IMF ($E_\mathrm{SN,Salpeter}$), such that
\begin{eqnarray}
    E_\mathrm{SN} &=& \eta ~ E_\mathrm{SN,Salpeter},
\end{eqnarray}
we yield the relation between the SFEs of a top-heavy and a Salpeter IMF.
\begin{eqnarray}
    \frac{f_\star^\mathrm{eff}}{f_\mathrm{\star, Salpeter}^\mathrm{eff}} &=& \frac{\min\left( \frac{v_c^2}{v_c^2+f_w E_\mathrm{SN}}, f_\star' \right)}{\min\left( \frac{v_c^2}{v_c^2+f_w E_\mathrm{SN,Salpeter}}, f_\star' \right)} 
    \label{eq_feff_relation} \\
    &\simeq&
    \begin{cases}
        1/\eta \ \ &\mathrm{for}\ \frac{f_w E_\mathrm{SN,Salpeter}}{v_c^2} \gtrsim 10\\
        [1/\eta, 1] \ \ &\mathrm{for}\ 10^{-3}\gtrsim \frac{f_w E_\mathrm{SN,Salpeter}}{v_c^2} \lesssim 10\\
        1 \ \ &\mathrm{for}\ \frac{f_w E_\mathrm{SN,Salpeter}}{v_c^2} \lesssim 10^{-3}
    \end{cases} \nonumber
\end{eqnarray}
In our top-heavy IMF parameterisation, $\eta$ increases from $\sim1$ to $\sim2.6~(3.4)$ as $f_\mathrm{massive}$ rises, assuming a maximum stellar mass for SNe explosions of $M_\mathrm{SN}^\mathrm{max}=50~(100)~\msun$. The relations between the SFEs for stars forming according to a top-heavy IMF and those forming according to a Salpeter IMF differ for low-mass (SN feedback-limited) and massive galaxies. To understand these different scalings and their impact on the UV enhancement for top-heavy IMFs, we show in Fig.~\ref{fig_fmassive_dependence} the halo mass dependence of the SFRs\footnote{$\mathrm{SFR}=f_\star^\mathrm{eff} \frac{\Omega_b}{\Omega_m} M_h$} (top panel, blue lines) and UV magnitudes (bottom panel, blue lines) of gas-rich galaxies for the Salpeter IMF (black) and increasingly top-heavier IMFs (dark to bright). The green lines in the top panel depict the factor by which stars forming during the past $30$~Myrs according to a top-heavy IMF are brighter than those forming according to a Salpeter IMF, accounting for the SFE relation in Eqn.~\ref{eq_feff_relation}. The assumed values are $f_w=0.2$ and $f_\star'=0.027$ based on $\Delta t=11$~Myr and $z=13$.

In lower-mass galaxies ($M_h\lesssim10^{9.5}\msun$), where star formation is limited by SN feedback, the SFE decreases for a top-heavy IMF by the same amount that its SN energy increases compared to a Salpeter IMF. As the IMF becomes more top-heavy, the SFE declines because stronger and more instantaneous SN feedback, driven by the presence of more massive stars with shorter lifetimes, suppresses subsequent star formation more quickly \citep[see also][]{Cueto2024}. This can be seen in the top panel of Fig.~\ref{fig_fmassive_dependence} where the SFR below $M_h\lesssim10^9\msun$ decreases with rising $f_\mathrm{massive}$ values. 
At the same time, the resulting UV luminosity per stellar mass that is formed nevertheless increases as $f_\mathrm{massive}$ rises (c.f. green lines). Yet, most of this UV luminosity enhancement occurs for values of $f_\mathrm{massive}\lesssim0.5$, up to a factor of $\sim3$, corresponding to a shift in $\Delta M_\mathrm{UV}\simeq-1.2$. Any further increase in $f_\mathrm{massive}$ enhances the UV luminosity by less than $10$\%. The reason for this decrease in UV luminosity enhancement is as follows: When $f_\mathrm{massive}>0.525$, the stellar mass at which the IMF slope changes, $M_c$, drops below the minimum stellar mass exploding as SNe, $M_\mathrm{SN,min}$; the mass distribution of the stars exploding as SNe is described by a log-flat IMF. Increasing $f_\mathrm{massive}$ above $0.525$ solely raises the normalisation of the IMF for $M>M_\mathrm{SN,min}$, resulting in a more significant increase in the number of SNe. However, the rise in UV luminosity with increasing $f_\mathrm{massive}$ remains relatively consistent at both lower and higher $f_\mathrm{massive}$ values, as contributions come from stars of all masses.
As a result, SN feedback-limited galaxies can only experience a UV luminosity boost up to $\Delta M_\mathrm{UV}=-1.3$, even with a fully log-flat IMF. However, due to the stronger SN feedback, a more top-heavy IMF results in more massive galaxies, with their star formation limited by SN feedback. As the break in the maximum intrinsic (i.e. dust unattenuated) UV luminosity, $M_\mathrm{UV,max}$, in the bottom panel in Fig.~\ref{fig_fmassive_dependence} indicates, at $z=13$, galaxies forming stars according to the Salpeter IMF and expelling all their star-forming gas have halo masses of $M_h\lesssim10^{8.8}\msun$, while this mass limit increases by $\sim0.8$~dex to $M_h\lesssim10^{9.6}\msun$ for a complete log-flat IMF.

In more massive galaxies ($M_h\gtrsim10^{9.5}\msun$), SN feedback is not strong enough to completely shut down star formation, leaving the SFE unaffected when moving from a Salpeter to a more top-heavy IMF (see the SFR above $M_h\simeq10^{9.5}\msun$ in the top panel of Fig.~\ref{fig_fmassive_dependence}). Given there is no decrease in the SFE as the IMF becomes more top-heavy (higher $f_\mathrm{massive}$ values), the UV luminosity enhancement due to the presence of more massive stars not only exceeds that in SN feedback-limited galaxies but also does not stall around $f_\mathrm{massive}\simeq0.5$; instead, it accelerates. This accelerating rise in UV luminosity enhancement is because replacing multiple lower-mass stars with massive stars increases the net UV luminosity due to the mass-luminosity relation of main-sequence stars where $L\propto M^{3.5}$. A log-flat IMF can boost the UV luminosity of more massive galaxies by a maximum of $\Delta M_\mathrm{UV}\simeq-2.6$ compared to a Salpeter IMF, which is twice as much as in an SN feedback-limited regime.

\section{The UV luminosity functions at $z=5-15$}
\label{sec_uvlfs}

\subsection{Model parameter dependence}
\label{subsec_model_parameter_dependence}

We find the free model parameters for the Salpeter and Evolving IMF models as follows. First, we estimate the SFE ($f_\star$) and SN wind coupling efficiency ($f_w$) that best reproduce the observed UV LFs at $z=5-10$ for the Salpeter IMF model (see Tab.~\ref{tab_astraeus_param} for values). Next, we adopt the same $f_\star$ and $f_w$ values for our Evolving IMF model, assume $z_\mathrm{max}=20$, and find the $z_\mathrm{thresh}$ and $k$ values that best reproduce the observed UV LF at $z=5-15$.
The free model parameters $z_\mathrm{thresh}$ and $k$ in our Evolving IMF model determine how $f_\mathrm{massive}$, describing the top-heaviness of the IMF, depends on the galaxies' gas content and sSFR values. 

Increasing $z_\mathrm{thresh}$ corresponds to decreasing the sSFR threshold ($\mathrm{sSFR_{thresh}}$) above which a galaxy's IMF becomes top-heavy. As a result, fewer galaxies will have a top-heavy IMF because their sSFR values increase with rising redshift due to their decreasing dynamic times, thus increasing SFEs. In the context of the UV LF, this reduction in the fraction of galaxies with top-heavy IMFs shifts the redshift at which the Evolving IMF model's UV LF starts exceeding that of the Salpeter IMF model to higher redshifts. For instance, for $z_\mathrm{thresh}=5$, we find the Evolving IMF model's UV LF to surpass the Salpeter IMF's UV LF at $z\gtrsim6$ and for $z_\mathrm{thresh}=6$ at $z\gtrsim7$ for large $k$ values ($>5$).
However, the exact redshift at which the Salpeter IMF model's UV LF is surpassed depends on the galaxies' actual gas-to-DM mass ratio and the value assumed for $k$, which regulates how top-heavy a galaxy's IMF becomes given its sSFR value or gas-to-DM mass ratio: the larger $k$ is, the more top-heavy is a galaxy's IMF for the same gas-to-DM ratio. In summary, while $z_\mathrm{thresh}$ determines how many galaxies exhibit a top-heavy IMF, $k$ regulates how top-heavy their IMFs are. 

Raising $k$ leads on average to higher $f_\mathrm{massive}$ values, more top-heavier IMFs and a boost of the UV luminosities, shifting the UV LF to higher luminosities. However, as discussed in Section~\ref{sec_UVboost}, the UV luminosity enhancement stalls for SN feedback-limited galaxies around $f_\mathrm{massive}\simeq0.5$ and hardly increases for values beyond. Thus, below a certain UV luminosity threshold, $M_\mathrm{UV,thresh}$, increasing $k$ beyond a value that corresponds to the majority of top-heavy galaxies having $f_\mathrm{massive}\gtrsim0.5$ values results in no significant further shift of the UV LF to higher luminosities. 
Beyond this UV luminosity threshold ($M_\mathrm{UV} < M_\mathrm{UV,thresh}$), the UV luminosities of top-heavy galaxies can be further increased by raising $f_\mathrm{massive}$ beyond 0.5. However, it is noteworthy that as $f_\mathrm{massive}$ increases, the halo mass and UV luminosity thresholds for galaxies to become SN feedback-limited also move to higher values, which is why we do not see abrupt changes. Given the different UV luminosity enhancement of SN feedback-limited and more massive galaxies, raising $k$ above a certain value leads to the bright end of the UV LF being shifted towards higher luminosities than the faint end. Yet, this shift is limited by the maximum shift possible in the UV, $\Delta M_\mathrm{UV}$, and the fraction of galaxies with top-heavy IMFs.

The maximum mass above which stars do not explode as energetic SNe but collapse directly into black holes also affects how much the UV luminosities of galaxies with top-heavy IMFs are boosted compared to a Salpeter IMF. Decreasing $M_\mathrm{SN}^\mathrm{max}$ reduces the SN feedback strength (or number of SNe explosions) fractionally more for a top-heavy than for a Salpeter IMF, since more massive stars directly collapse into black holes. This decrease in SN feedback causes the galaxies with top-heavy IMFs to exhibit higher UV luminosities, shifting the UV LFs to higher luminosities. 
We find that, for $M_\mathrm{SN}^\mathrm{max}=100~\msun$, the SN feedback is essentially too strong to boost our Evolving IMF model's UV LFs to the observational data points. Only decreasing $M_\mathrm{SN}^\mathrm{max}$ to the more realistic value of $\lesssim50~\msun$ allows a reasonable fit. 
Given the trends described above, for our Evolving IMF, we choose $M_\mathrm{SN}^\mathrm{max}=50~\msun$, which is supported by the SNe failure rate of $>30~\msun$ stars \citep[e.g.][]{Kobayashi2020b}, and find $z_\mathrm{thresh}=6$ and $k=7$ to best-fit the observed UV LF at $z=5-15$. These best-fit values have been determined by comparing the simulated dust-attenuated UV LFs to the observational data. We note, due to our coarse gridding of $k$ and $z_\mathrm{thresh}$ (with $\Delta k=1$ and $\Delta z_\mathrm{thresh}=1$), the best-fit values have been identified visually. Additionally, we find that both IMF models' stellar mass functions (SMFs) are consistent with the values inferred from observations \citet{song_evolution_2016, duncan_mass_2014, gonzalez_evolution_2011}.

\begin{table}
    \centering
    \caption{Best fit {\sc astraeus} model parameters for Salpeter and Evolving IMFs}
    \label{tab_astraeus_param}
    \begin{tabular*}{\columnwidth}{cccc}
        \hline
        Parameter & Salpeter & Evolving & Description \\
        \hline
        \hline
        $f_\star$ & 0.03 & 0.03 & SFE normalisation\\
        $f_w$ & 0.2 & 0.2 & wind coupling efficiency \\
        IMF & Salpeter & Evolving & IMF \\
        $z_\mathrm{max}$ & - & 20 & IMF parameter\\
        $z_\mathrm{thresh}$ & - & 6 & IMF parameter\\
        $k$ & - & 7 & IMF parameter\\
        \hline
    \end{tabular*}
\end{table}

\subsection{Best-fit case}
\label{subsec_bestfit_case}

\begin{figure*}
\resizebox{\hsize}{!}
        {\includegraphics{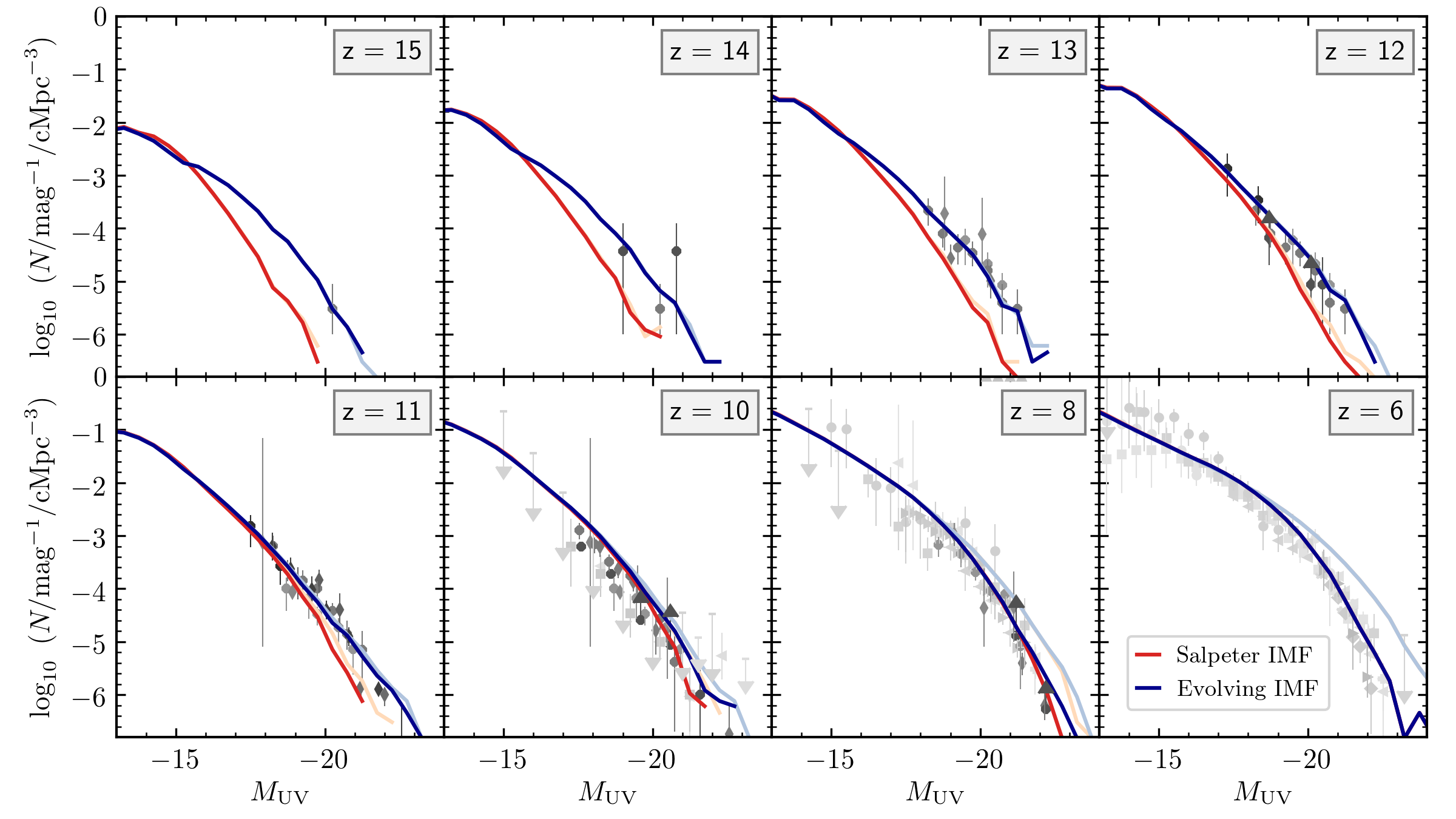}}
    \caption{Ultraviolet luminosity functions for the Salpeter IMF (red lines) and Evolving IMF (blue lines) at $z=6-15$ including (dark coloured lines) and without (light coloured lines) dust attenuation. Light grey points depict results from observations with the Hubble Space Telescope \citep{Atek2015, Atek2018, Bouwens2021, Castellano2010, Ishigaki2018, Livermore2017, McLeod2015, McLeod2016, McLure2013, Oesch2018, Schenker2013}, while darker grey points show the results from JWST observations \citep{Adams2024, Bouwens2023a, Bouwens2023b, Donnan2023a, Donnan2023b, Donnan2024, Finkelstein2023, McLeod2024, Harikane2024, PerezGonzalez2023}.}
    \label{fig_uvlfs}
\end{figure*}

Next, we discuss the UV LFs for our best-fit parameter set and the physical processes that drive their shape and evolution at $z>10$. Fig.~\ref{fig_uvlfs} shows the intrinsic (light) and observed (dark) UV luminosity functions (LFs) for the Salpeter IMF (red lines) and Evolving IMF (blue lines) models at $z=6$, $8$, $10-15$. 

The UV LFs of both IMF models reflect the physical processes driving their redshift evolution. As redshift decreases, their shift to higher UV luminosities and number densities reflects hierarchical structure formation, with galaxies growing in mass through mergers and accretion and increasingly more low-mass galaxies forming, raising the luminosity of the brightest galaxies (from $M_\mathrm{UV}\simeq-20$ and $-22$ for the Salpeter and Evolving IMF models at $z=15$ to $-24$ at $z=6$) and the number of UV-faint galaxies (from $\sim10^{-2.5}$~cMpc$^{-3}$ at $z=15$ to $\sim10^{-3}$~cMpc$^{-1.4}$ at $z=6$ for $M_\mathrm{UV}\simeq-16$), respectively. Despite their rising numbers, lower-mass galaxies not only show bursty star formation \citep[e.g.][]{Legrand2022} due to SN feedback and radiative feedback from reionisation -- leading to a broadening of their UV luminosity values -- but are also consumed by mergers in the vicinity of more massive galaxies. Both govern the UV LF's number density and luminosity evolution, causing the flattening of its slope at the faint end and developing a break around $M_\mathrm{UV}\simeq-18$ towards lower redshifts. 

As can be seen from Fig.~\ref{fig_uvlfs}, both IMF models, Salpeter and Evolving IMF, reproduce the observed UV LFs at $z\lesssim10$, capturing both the normalisation and the knee that develops around $M_\mathrm{UV}\simeq -18$ at lower redshifts. However, at higher redshifts ($z>10$), the Salpeter IMF model underpredicts the observed UV LFs, particularly at the bright end ($M_\mathrm{UV}\lesssim-19$), while the Evolving IMF reproduces the observed UV LF at $z=10-15$. 

The reasons for this better fit are as follows: Firstly, towards higher redshifts, a rising fraction of galaxies exhibit top-heavy IMFs, leading to an increasing shift towards higher UV luminosities compared to the Salpeter IMF. For example, at $z=11$, where only $\sim0.8\%$ of the simulated galaxies ($M_h\geq10^{8.3}\msun$) have a top-heavy IMF, the UV LF shifts only at the bright end by less than $1$~dex towards higher luminosities. However, at $z=15$, the fraction of galaxies with top-heavy IMFs increases to $\sim8\%$, resulting in a shift of the UV LF by about $1$~dex.
Secondly, the range of IMFs of the galaxies with top-heavy IMFs reaches top-heavier IMFs, i.e. larger $f_\mathrm{massive}$ values (c.f. Eqn.~\ref{eq_fmassive}), as redshift increases, enabling larger shifts towards brighter UV luminosities even for SN-feedback limited galaxies.
At $z=6$, the UV LF matches that of the Salpeter IMF model, since we assume no galaxies have a top-heavy IMF at this redshift ($f_\mathrm{massive}=0$).

A key difference between the UV LFs of the Evolving and Salpeter IMF models at $z>10$ are the shallower slopes observed in the Evolving IMF model. This difference arises for two reasons: Firstly, the average gas fraction, and thus the fraction of galaxies with a top-heavy IMF, increases with halo mass \citep[see e.g.][]{Hutter2021a}, leading to fewer low-mass and faint galaxies having a top-heavy IMF than at the massive and bright end. Secondly, as shown in Fig.~\ref{fig_fmassive_dependence} and discussed in Section~\ref{sec_UVboost}, SN feedback-limited and UV-faint galaxies experience a smaller UV luminosity enhancement for a top-heavy IMF with the same $f_\mathrm{massive}$ value than more massive and UV brighter galaxies. 
Both the higher fraction of UV-bright galaxies with a top-heavy IMF and the more significant UV luminosity enhancement in these galaxies cause the bright end of the UV LF to shift more towards higher UV luminosities than the faint end. 

It is also important to note that UV-bright galaxies ($M_\mathrm{UV}\leq -18$) in the Evolving IMF model fall into two categories: those that are bright due to their accumulated stellar mass and those that are bright because they form stars according to a top-heavy IMF. The former have stellar and dust masses comparable to similarly massive galaxies in the Salpeter IMF model,\footnote{However, as the fraction of galaxies with top-heavy IMFs increases towards higher redshifts, the stellar-to-halo mass and dust-to-halo mass ratios decrease slightly in massive galaxies, while the dust-to-stellar mass ratio remains unchanged.} corresponding to $M_\star\simeq10^{7.3-8.3}\msun$ and $M_\mathrm{d}\simeq10^{4-5}\msun$ at $z=13$, respectively. In contrast, the latter are less massive galaxies ($M_\star\simeq10^{6.1-7.3}\msun$ at $z=13$) whose UV luminosities are boosted by the higher fraction of massive stars. These galaxies have lower dust masses ($M_\mathrm{d}\simeq10^{2-3.5}\msun$ at $z=13$), which reduces their UV dust attenuation and results in a smaller difference between the intrinsic (bright blue lines) and observed (dark blue lines) UV LFs in Fig.~\ref{fig_uvlfs}.

Finally, we briefly discuss why our results differ from those in \citet{Cueto2024}. In \citet{Cueto2024}, the $f_\mathrm{massive}$ values exceed $\sim0.2$ and increase towards higher redshifts and lower gas-phase metallicities. Due to the dependence on gas-phase metallicity, low-mass, SN feedback-limited galaxies generally have larger $f_\mathrm{massive}$ values than more massive galaxies. When all galaxies form stars according to a top-heavy IMF, the decreasing $f_\mathrm{massive}$ with increasing galaxy mass and gas-phase metallicity results in similar UV luminosity enhancements for both SN feedback-limited and more massive galaxies (see the top panel of Fig.~\ref{fig_fmassive_dependence}). Consequently, assuming the same $f_\star$ and $f_w$ values as for the Salpeter IMF would shift the UV LFs to higher luminosities by approximately $\sim0.5$~dex for massive galaxies and $\sim1$~dex for SN feedback-limited galaxies. However, \citet{Cueto2024} compensates for this shift by decreasing the SFE normalisation by factors of $\sim0.4$ and $\sim0.7$ for massive and SN feedback-limited galaxies, respectively (corresponding to $\Delta M_\mathrm{UV} \simeq 0.5$ and $1$), to match the observed UV LFs at $z=5-13$. 

\section{The galaxy population with top-heavy IMFs}
\label{sec_galpop}

\begin{figure}
    \centering
    \includegraphics[width=\hsize]{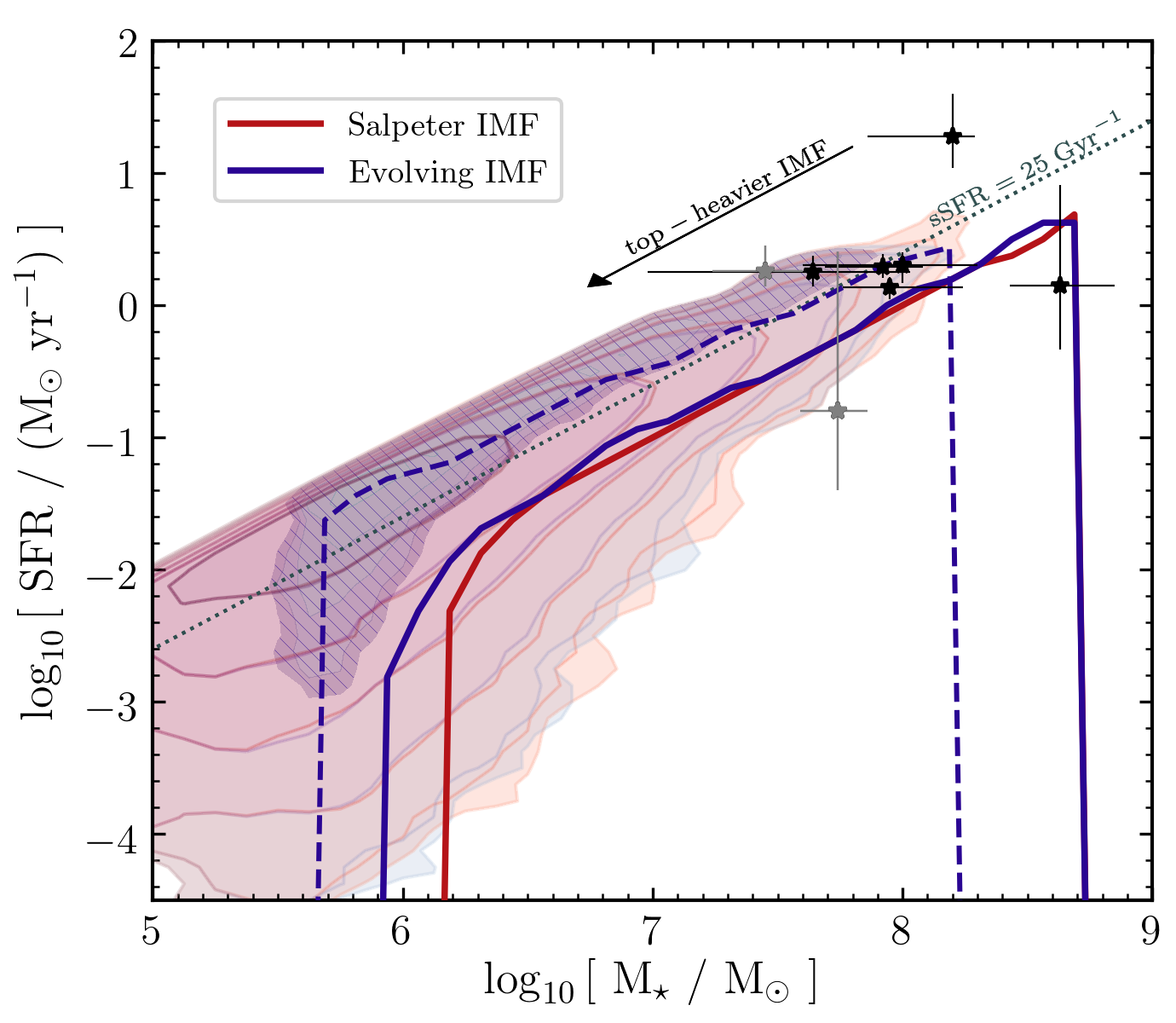}
    \caption{SFR - stellar mass relation for the Salpeter IMF (red) and the Evolving IMF (blue) at $z=13$. Hashed blue contours show the galaxies with an IMF top-heavier than the Salpeter IMF in the Evolving IMF model. Black points show the inferred values from JADES-GS-z14-1, JADES-GS-z13-0, JADES-GS-z12-0, UNCOVER-z13, UNCOVER-z12, and GLASS-z12 \citep{CurtisLake2023, Wang2023, Bakx2023}. grey points show the two set of values inferred for JADES-z13-1-LA, where assuming no nebular emission leads to a higher SFR and lower stellar mass \citep{Witstok2024}.}
    \label{fig_sSFR}
\end{figure} 

To complete the picture, we analyse which galaxies within the Evolving IMF model exhibit a top-heavy IMF. As the IMF's top-heaviness -- defined as the fraction of mass produced in the log-flat part of the IMF -- scales with the gas-to-DM mass ratio according to Eqn.~\ref{eq_fmassive}, we find that galaxies with top-heavy IMFs are typically gas-rich and highly star-forming, exceeding a threshold of $\mathrm{sSFR}\simeq25$~Gyr$^{-1}$. This is illustrated in Fig.~\ref{fig_sSFR}, which shows the SFR-stellar mass relations for the Evolving IMF (blue contours) and Salpeter IMF (red contours) at $z=13$, along with the SFR - $M_\star$ combinations for galaxies with top-heavy IMFs ($f_\mathrm{massive}>0$, blue hashed contours). Indeed, the median SFRs of the galaxies with top-heavy IMFs (dashed blue lines) exceed the median SFR of the entire galaxy population (solid blue lines) by $\sim0.5$~dex. The sSFR threshold of $25$~Gyr$^{-1}$ is marked by the grey dotted line.

However, from the blue hashed contours in Fig.~\ref{fig_sSFR}, we note two deviations from the aforementioned selection criterion for galaxies with top-heavy IMFs, both arising from the limited validity of the assumption that the sSFR scales with the gas-to-DM mass ratio, as assumed in Eqn.~\ref{eq_fmassive}.
Firstly, in galaxies with lower stellar masses ($M_\star\lesssim10^{6.5}~\msun$), increasingly lower sSFR values also exhibit top-heavy IMFs, i.e. the blue hashed contours extend below the sSFR threshold (grey dotted line). In these lower-mass galaxies, star formation is increasingly suppressed by SN feedback and thus not primarily shaped by the gas accretion rate, resulting in a decrease in the lower sSFR limit.\footnote{In our delayed SN feedback scheme \citep[see][]{Hutter2021a}, the specific time delay between star formation and SN explosion is accounted for all stellar masses. Thus, even in short time steps, star formation can be suppressed due to SNe from massive stars formed in the previous time step.}
Secondly, low-mass galaxies with $M_\star\lesssim10^6~\msun$ exhibit no top-heavy IMFs despite some of them exceeding a sSFR of $25$~Gyr$^{-1}$. Below this mass threshold (corresponding to $M_h\simeq10^{8.7}~\msun$), the galaxies' SFHs have not entirely converged: these galaxies have existed for only a few time steps and thus could not build up the stellar mass they would have if the merger trees extended to lower halo masses than $M_h=10^{8.2}~\msun$.
Both deviations highlight that a better approximation for the sSFR value of SN feedback-limited galaxies would be desirable when determining whether a galaxy exhibits a top-heavy IMF. However, this is challenging if not impossible since the SFR depends on the assumed IMF. 

Galaxies with high sSFRs are typically gas-rich, sustained by high gas accretion rates that replenish their gas reservoirs and counteract the losses due to SNe. These galaxies are usually found in environments where they are the deepest potential wells or most massive galaxies \citep[see e.g.][]{Legrand2023}. Consequently, low-mass galaxies with top-heavy IMFs are located in slightly above-average dense regions, while their massive counterparts are found in overdense regions, resulting in their median environmental density being slightly above that of the overall galaxy population. Furthermore, due to their higher gas accretion rates, which continuously dilute the ongoing metal enrichment, galaxies with top-heavy IMFs have the lowest gas-phase metallicities (oxygen-to-hydrogen mass ratio) despite their top-heavy IMFs producing in their SNe explosions more oxygen per stellar mass.

How does the population of galaxies with top-heavy IMFs evolve? As the matter density in the Universe decreases with cosmic time, the time for matter to collapse and cool in halos of a given mass lengthens, reducing the available cool gas to form stars. Consequently, the sSFRs of galaxies with a given mass decrease over time, and the SFR-$M_\star$ relation shifts to lower SFR values. Since our selection of galaxies with top-heavy IMFs effectively corresponds to a cut-off at an sSFR of $25$~Gyr$^{-1}$, fewer galaxies exhibit a top-heavy IMF at lower redshifts, and those that do increasingly approach the lower threshold of sSFR$_\mathrm{thresh} \simeq 25$~Gyr$^{-1}$, similar to the $f_\mathrm{massive}$ values described in Section~\ref{subsec_model_parameter_dependence}. While our simulated galaxies show sSFR values up to $130$~Gyr$^{-1}$ at $z=15$, their sSFR only reaches $80$~Gyr$^{-1}$ at $z=10$. 
Interestingly, the sSFR threshold inferred and required to reproduce the UV LFs at $z=5-15$ aligns closely with the sSFR value identified by \citet{Ferrara2023}, above which radiation pressure on dust is sufficient to expel it from the star formation sites. Although our predictions for the UV LFs at $z>10$ also assume negligible dust attenuation, only the Evolving IMF model successfully reproduces the observed number densities.

Finally, we note that assuming more top-heavy IMFs does not significantly alter the simulated SFR-$M_\star$ relation. As explained in \citet{Cueto2024}, this insensitivity of the star formation sequence arises from the self-similar mass growth of halos and, consequently, the gas and stellar mass. Making the IMF more top-heavy reduces the SFR and stellar mass, shifting the galaxy along the lines of constant sSFR, as illustrated by the black arrow in Fig.~\ref{fig_sSFR}.
This characteristic becomes important when comparing our predictions to observed galaxy properties. Since most observational data analysis pipelines assume Kroupa, Chabrier, or Salpeter IMFs, the stellar masses and SFRs inferred for galaxies with top-heavy IMFs are likely overestimated. These galaxies would then appear to shift along lines of constant sSFR to lower stellar mass and SFR values, which importantly does not affect our criterion for identifying top-heavy IMFs.\footnote{Here we do not consider potential biases in SFR and stellar mass estimates that may arise from deriving these quantities from different wavelength ranges of galaxy spectra or emission lines.} For this reason and given the complexity of the inference of galaxy properties from observations, we omit to adjust the inferred values from observations to our IMFs in Fig.~\ref{fig_sSFR}.

\paragraph{Comparison with observations:}

In Fig.~\ref{fig_sSFR}, we mark with black points where JADES-GS-z14-1, JADES-GS-z13-0 and JADES-GS-z12-0 \citep{CurtisLake2023}, UNCOVER-z13 and UNCOVER-z12 \citep{Wang2023} as well as GLASS-z12 \citep{Bakx2023} fall within the star formation main sequence of our simulations. Grey points represent the inferred values for JADES-z13-1-LA, both for their default model and an alternative model without nebular emission, with the latter leading to higher SFR and lower stellar mass values.
Except for GLASS-z12, the only galaxies with SFR inferred from ALMA observations, JADES-GS-z12-0 is the only galaxy that has an SFR-$M_\star$ combination clearly indicative of a top-heavy in our Evolving IMF model. The observed spectrum exhibiting no strong emission lines indeed suggests a high ionising photon production efficiency of $\log_{10}(\xi_\mathrm{ion}/\mathrm{erg}^{-1}\mathrm{Hz})=25.7$, close to the for typical ISM conditions and normal stellar populations of $25.8$, and a red UV slope of $\beta=-1.84$, possibly an indication of hot stars and a low-density ISM \citep{Katz2024}. Moreover, its low metallicity value of $\log_{10}(Z/Z_\odot)=-1.4$ also agrees with our model predictions. While the SFR-$M_\star$ combinations for JADES-z13-1-LA are inconclusive regarding a top-heavy IMF, its high ionising photon production efficiency ($\log_{10}(\xi_\mathrm{ion}/\mathrm{erg}^{-1}\mathrm{Hz})=26.5$) and extremely low metallicity ($\log_{10}(Z/Z_\odot)=-2.5$) strongly suggest the presence of such an IMF. 
Within uncertainties, JADES-GS-z14-1 and UNCOVER-z13 could also be candidates containing stellar populations with top-heavy IMFs given their inferred high SFRs of $\sim2~\msun/\mathrm{yr}$ but have metallicities at the upper end of what our simulated galaxies with top-heavy IMFs suggest.\footnote{JADES-GS-z14-1 shows no emission lines, indicating either very low metallicity or high ionising escape fractions, and its very blue UV slope ($\beta=-2.7$) points to a young stellar population with low dust attenuation.} However, UNCOVER-z12 and JADES-GS-z13-0 fall not into the range of galaxies with top-heavy IMFs. JADES-GS-z13-0, for instance, has an average UV slope of $\beta=-2.4$ and an ionising photon production efficiency of $\log_{10}(\xi_\mathrm{ion}/\mathrm{erg}^{-1}\mathrm{Hz})=25.5$, not necessarily pointing towards the extreme conditions associated with a top-heavy IMF.

Finally, we note that at $z=11$, GN-z11 \citep{Oesch2016, Bunker2023} also falls within the SFR-$M_\star$ combinations indicative of a top-heavy IMF in our Evolving IMF model. This hypothesis of a top-heavy IMF also aligns with further observational evidence, including the elevated [N/O] abundance, the increased ionising emissivity ($\xi_\mathrm{ion}=25.7$) and blue UV slope ($\beta=-2.4$) \citep[see e.g.][]{Cameron2024, Vink2023, Senchyna2024, Nandal2024}.

\section{The UV luminosity scatter for top-heavy IMFs}
\label{sec_UVlum_scatter}

\begin{figure}
    \centering
    \includegraphics[width=\hsize]{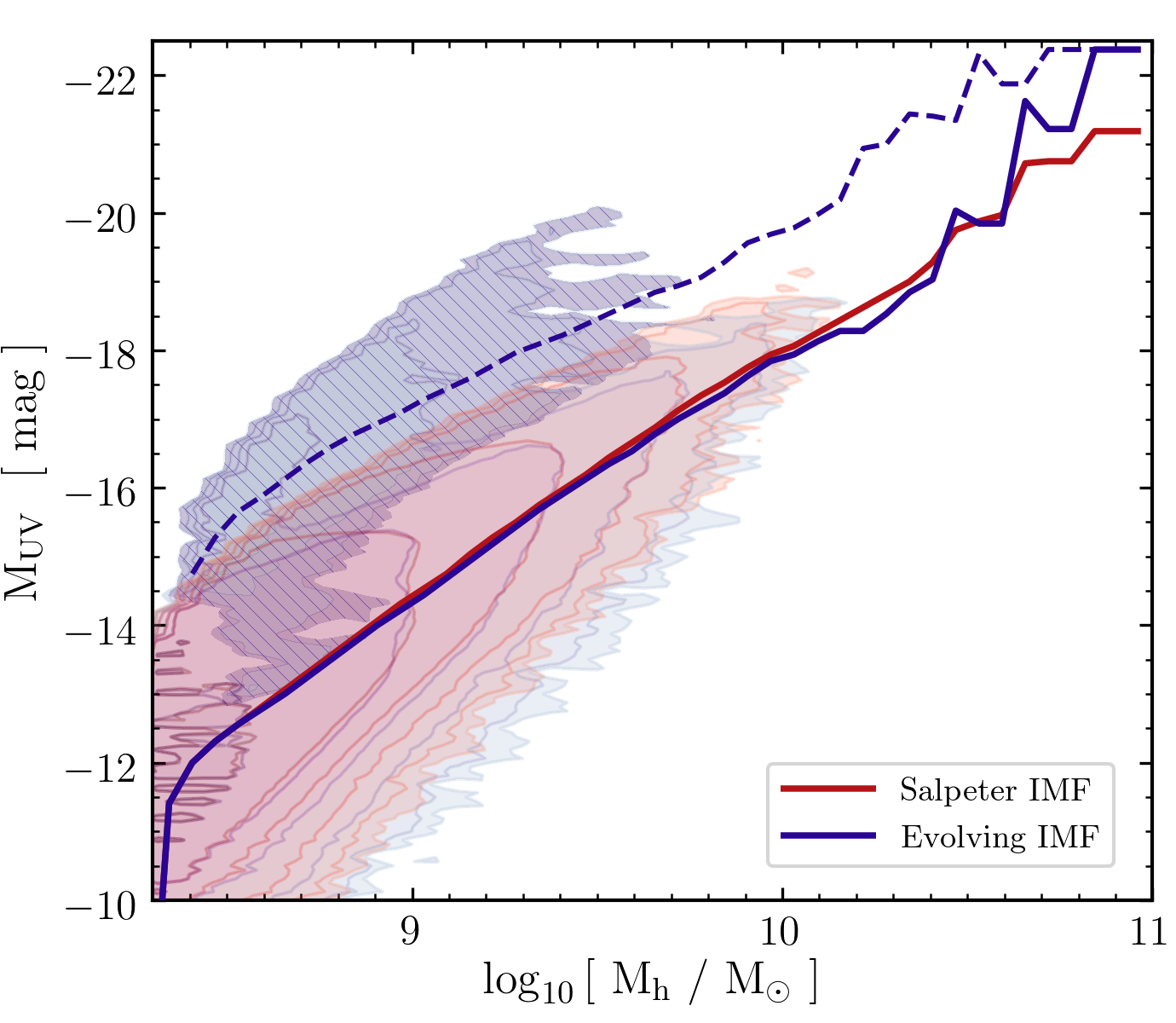}
    \caption{$M_\mathrm{UV}$ - $M_h$ relation for the Salpeter IMF (red) and the Evolving IMF (blue) at $z=13$. Hashed blue contours show the galaxies with an IMF top-heavier than the Salpeter IMF in the Evolving IMF model.}
    \label{fig_MUV-Mvir}
\end{figure} 

\begin{figure}
    \centering
    \includegraphics[width=\hsize]{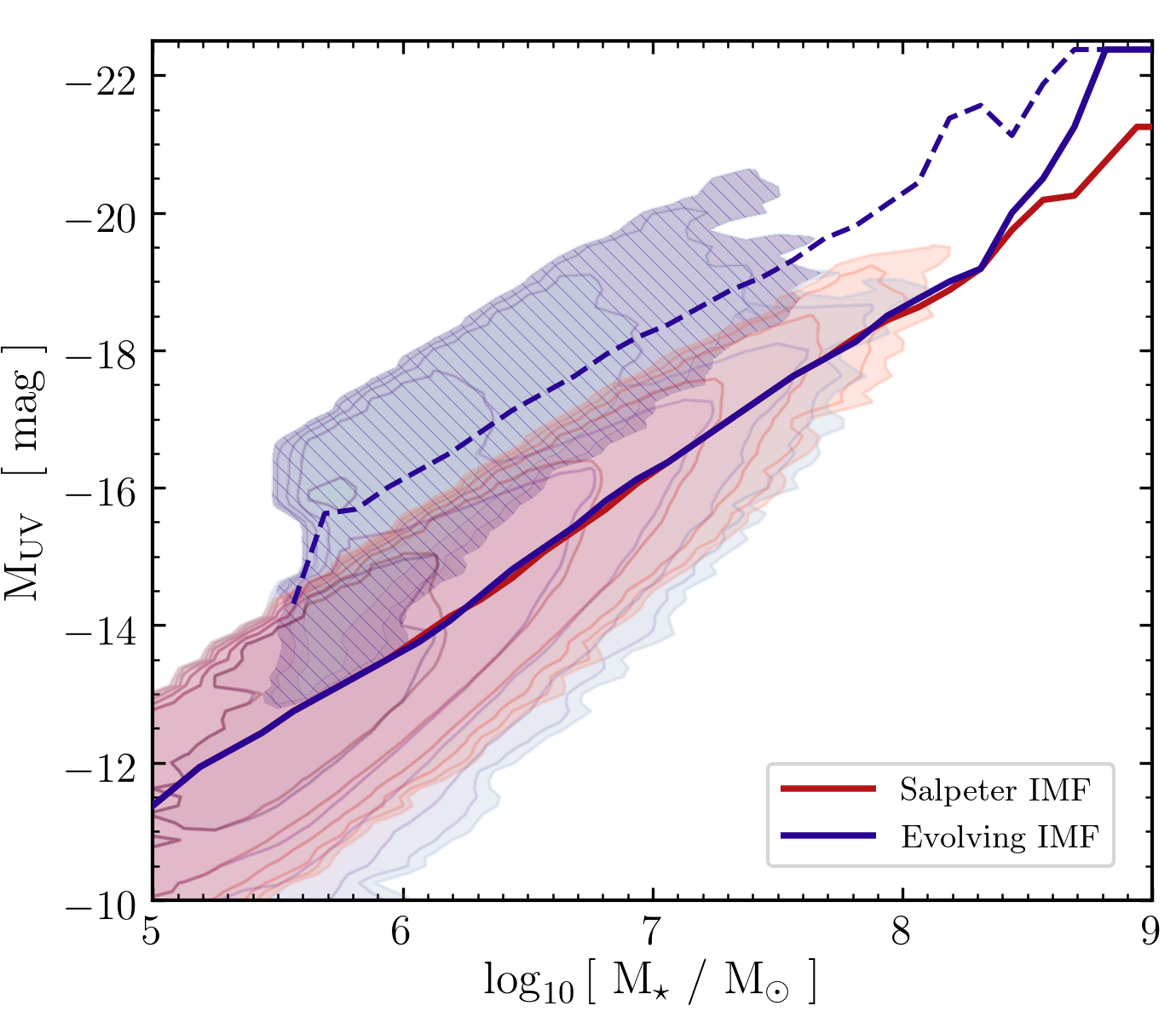}
    \caption{$M_\mathrm{UV}$ - $M_\star$ relation for the Salpeter IMF (red) and the Evolving IMF (blue) at $z=13$. Hashed blue contours show the galaxies with an IMF top-heavier than the Salpeter IMF in the Evolving IMF model.}
    \label{fig_MUV-Mstar}
\end{figure} 

Section~\ref{sec_UVboost} discussed how the maximum possible UV luminosity enhancement when transitioning from a Salpeter to a top-heavy IMF varies with halo mass. Fig.~\ref{fig_MUV-Mvir} shows the relation between UV luminosity ($M_\mathrm{UV}$) and halo mass ($M_h$) for our Salpeter IMF (red contours) and Evolving IMF model (blue contours) simulations. Consistent with our toy model, we find that simulation-based star formation and gas mass histories also lead to a UV luminosity enhancement of up to $\sim1$~dex for SN feedback-limited galaxies ($M_h\lesssim10^{9}\msun$). This enhancement increases towards more massive galaxies, reaching up to $\sim2.5$~dex, as the transition from the SN feedback-limited to constant SFE regime shifts to higher masses with rising $f_\mathrm{massive}$ values. 
Additionally, the Evolving IMF model produces by up to $\sim0.5$~dex fainter galaxies at the same halo mass. Stronger and more immediate SN feedback in galaxies with top-heavy IMFs suppresses star formation more quickly than in galaxies with a Salpeter IMF due to more massive stars with shorter lifetimes \citep[see also][]{Cueto2024}, making these galaxies fainter in the UV. 

Tables~\ref{tab_UVlum_scatter_Salpeter} and \ref{tab_UVlum_scatter_Evolving} present the median $M_\mathrm{UV}$ values along with their $1\sigma$ and $2\sigma$ uncertainties across various halo masses for the Salpeter and Evolving IMF models, respectively. While the median $M_\mathrm{UV}-M_h$ relations hardly differ between the two IMF models, the Evolving IMF model exhibits a larger variance in UV luminosity than the Salpeter IMF model, with $1\sigma_\mathrm{UV}$ and $2\sigma_\mathrm{UV}$ values being by about a factor $1.1-1.3$ and $1.4-2$ larger. Although the $\sigma_\mathrm{UV}$ values in both IMF models are lower than those inferred from the FIRE simulations \citep[$\sigma_\mathrm{UV}\sim1.6$ to $0.9$ from $M_h=10^{8.5}\msun$ to $10^{10.5}\msun$;][]{Sun2023} or noted as necessary to reproduce the $z>10$ UV LFs in analytic models \citep[$\sigma_\mathrm{UV}\simeq2$ at $z=13$;][]{Mason2023, Munoz2023, Shen2023, Kravtsov2024}, the boost in the $2\sigma$ variances when transitioning from the Salpeter to the Evolving IMF model are significantly larger and align better with the variances found in the aforementioned analytic models reproducing the $z>10$ UV LFs. 
This indicates that the UV luminosity distribution is stretched more at the extreme values ($>1\sigma$) rather than near the median ($<1\sigma$) - a result stemming from the assumption that only the most gas-rich galaxies in relatively denser regions exhibit top-heavy IMFs and thus enhanced UV luminosities. While this boost in $>1\sigma$ values in clustered locations may enhance the clustering signal, the increased variance may also lead to a decrease in the clustering signal as shown in \citet{Gelli2024}. In future work, we will investigate whether the Evolving IMF leaves a characteristic imprint in large-scale clustering.

\begin{table}
    \centering
    \caption{UV luminosity median and variance values for the Salpeter IMF}
    \label{tab_UVlum_scatter_Salpeter}
    \begin{tabular*}{0.6\columnwidth}{cccc}
        \hline
        $\log_{10} M_h/\msun$ & $M_\mathrm{UV}$ & $\sigma_\mathrm{UV}$ & $2\sigma_\mathrm{UV}$ \\
        \hline
        \hline
        8.5 & -12.7 & 1.22 & 5.7 \\
        9   & -14.6 & 0.91 & 1.71 \\
        9.5 & -16.6 & 0.84 & 1.59 \\
        10  & -18.1 & 0.78 & 1.34 \\
        10.5 & -19.7 & 0.97 & 0.97 \\
        \hline
    \end{tabular*}
\end{table}

\begin{table}
    \centering
    \caption{UV luminosity median and variance values for the Evolving IMF}
    \label{tab_UVlum_scatter_Evolving}
    \begin{tabular*}{0.6\columnwidth}{cccc}
        \hline
        $\log_{10} M_h/\msun$ & $M_\mathrm{UV}$ & $\sigma_\mathrm{UV}$ & $2\sigma_\mathrm{UV}$ \\
        \hline
        \hline
        8.5 & -12.7 & 1.25 & 6.2 \\
        9   & -14.6 & 0.97 & 2.47 \\
        9.5 & -16.4 & 0.91 & 2.56 \\
        10  & -18.0 & 1.00 & 2.63 \\
        10.5 & -19.7 & 1.94 & 1.94 \\
        \hline
    \end{tabular*}
\end{table}

Fig.~\ref{fig_MUV-Mstar} shows the relation between the UV luminosity ($M_\mathrm{UV}$) and stellar mass ($M_\star$). Across all stellar masses where galaxies with top-heavy IMFs exist, we find that the maximum UV enhancement matches the $\Delta M_\mathrm{UV}$ of $\sim2.5$~dex derived from our toy model. This contrasts with the UV luminosity-halo mass relation, where the enhancement increases with rising halo mass. Since the UV luminosity strongly depends on the SFR, this difference arises because the SFR-$M_\star$ relation remains largely unchanged when top-heavy IMFs are included due to the self-similar growth of galaxy mass \citep[see Fig.\ref{fig_sSFR} and][]{Cueto2024}.

\section{Implications for reionisation}
\label{sec_reionisation}

\begin{figure}
    \centering
    \includegraphics[width=\hsize]{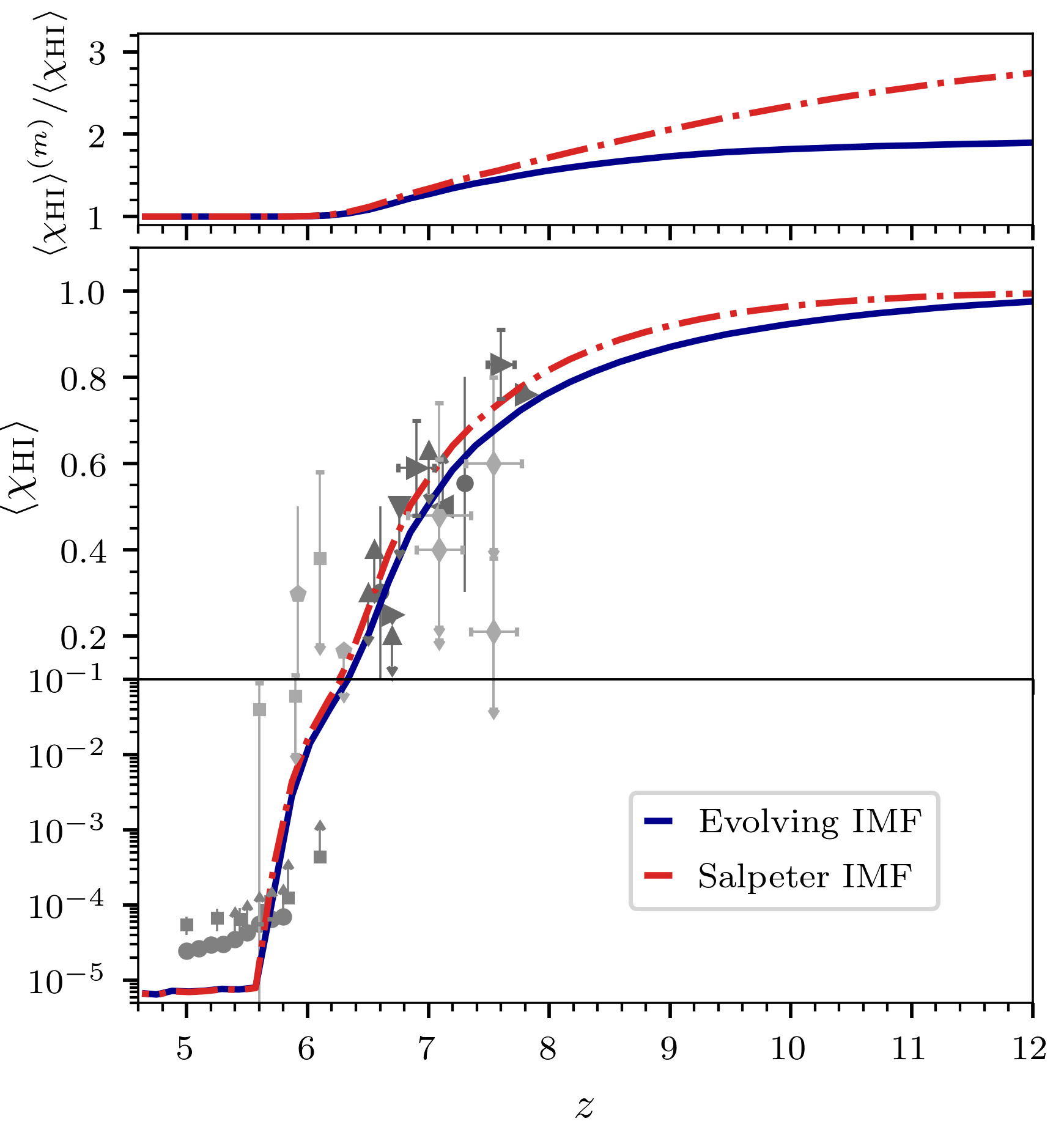}
    \caption{Redshift evolution of the global IGM \HI fraction for the Salpeter IMF (red) and Evolving IMF (blue) model (bottom) and the ratio between the mass-averaged and volume-averaged IGM \HI fractions (top). Grey points indicate observational constraints from: GRB optical afterglow spectrum analyses \citep[light pentagons;][]{Totani2006, Totani2014}, quasar sightlines \citep[faint squares;][]{Fan2006}, \citep[medium bright circles;][]{Bosman2022} Lyman-$\alpha$ LFs \citep[dark upward triangles;]{Konno2018,Kashikawa2011,Ouchi2010,Ota2010,Malhotra2004}, Lyman-$\alpha$ emitter clustering \citep[dark downwards triangles;][]{Ouchi2010}, the Lyman-$\alpha$ emitting galaxy fraction \citep[dark leftwards triangles;][]{Pentericci2011, Schenker2012, Ono2012, Treu2012, Caruana2012, Caruana2014, Pentericci2014}, Lyman-$\alpha$ equivalent widths \citep[dark rightwards triangles;][]{Mason2018a, Mason2019, Bolan2022}, dark pixels \citep[light squares;][]{McGreer2015}, and damping wings \citep[light diamonds;][]{Davies2018, Greig2019}.}
    \label{fig_ionisation_history}
\end{figure} 

The history and morphology of the reionisation of the neutral hydrogen in the IGM are highly sensitive to the ionising photon output and distribution of the first galaxies. We examine the reionisation process for both IMF models, assuming that the fraction of ionising photons escaping from a galaxy into the IGM scales with the fraction of gas ejected from the galaxy, expressed as $f_\mathrm{esc}=f_\mathrm{esc}^0 \min(1, f_\star^\mathrm{eff} / f_\star^\mathrm{ej})$ \citep[see e.g.][]{Hutter2021a, Hutter2023a}, with $f_\mathrm{esc}^0=0.35$. The corresponding reionisation histories, the redshift evolution of the global IGM \HI neutral fraction, $\langle \chi_\mathrm{HI}\rangle$, are shown in Fig.~\ref{fig_ionisation_history}.

Raising the fraction of massive stars increases the ionising emissivity of stellar populations by up to an order of magnitude, similar to the UV luminosity. Consequently, as more galaxies form stars according to a top-heavy IMF at high redshift, the Evolving IMF model (blue line) results in an earlier onset of reionisation ($\Delta z\simeq2$ at $\langle \chi_\mathrm{HI} \rangle \simeq 0.99$) compared to the Salpeter IMF model (red line), characterised by ionised regions growing earlier around the massive galaxies with top-heavy IMFs. At the same global IGM neutral fractions, $\langle \chi_\mathrm{HI}\rangle$, the ionisation morphology in the Evolving IMF model shows more extended large ionised regions and fewer small ones. For example, at $z\simeq10$, the largest ionised regions in the Evolving IMF model extend up to $\sim10$~cMpc, while in the Salpeter IMF model, they barely reach $5$~cMpc.
As the fraction of galaxies with top-heavy IMFs decreases, the differences in $\langle\chi_\mathrm{HI}\rangle$ and ionisation morphology diminish: By $z\simeq7$, the ionisation morphologies become very similar, and the redshift difference at the same $\langle\chi_\mathrm{HI}\rangle$ values reduces to $0.2$. In both IMF models, reionisation completes by $z\simeq5.6$, consistent with constraints from QSO sightlines \citep[e.g.][]{Bosman2022}. 
Due to an earlier start of reionisation in the Evolving IMF model, the electron optical depth is slightly higher ($\sigma=0.055$) compared to the Salpeter IMF model ($\sigma=0.052$), but remains within the limits set by the Cosmic Microwave Background measurements with Planck \citep{planck2018}. A more detailed analysis of the differences in ionisation morphology will be presented in future work.

\section{Observational signatures of a top-heavy IMF}
\label{sec_xion}

\begin{figure}
    \centering
    \includegraphics[width=\hsize]{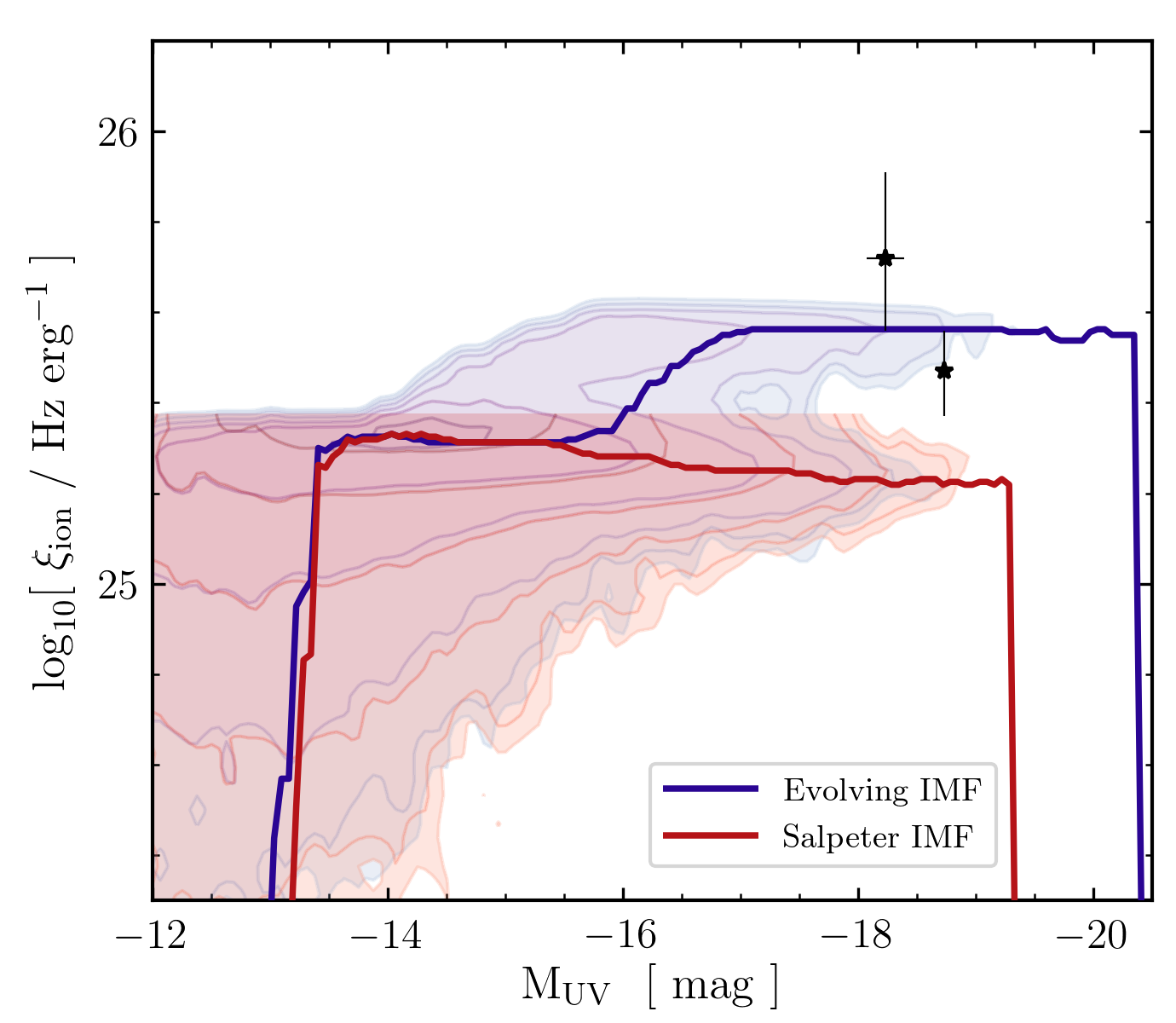}
    \caption{$\xi_\mathrm{ion}$ - $M_\mathrm{UV}$ relation for the Salpeter IMF (red) and the Evolving IMF (blue) at $z=13$. Solid lines show the median of the corresponding coloured contours. Black points show the inferred values for JADES-GS-z13-0 and JADES-GS-z12-0 \citep{CurtisLake2023}, with JADES-GS-z12-0 showing higher $\xi_\mathrm{ion}$ values that agree more with the Evolving IMF than Salpeter IMF model.}
    \label{fig_xiion-MUV}
\end{figure} 

Given the crucial role of the IMF in determining the ionising photon output — and consequently the reionisation history and morphology — we briefly discuss how the ionising emissivity, $\xi_\mathrm{ion} = \dot{Q}/L_\mathrm{UV}^\mathrm{int}$, differs at $z\gtrsim10$ in our two IMF models. $\xi_\mathrm{ion}$ has been inferred from galaxy observations by comparing the luminosity of emission lines (e.g. H$\alpha$, OIII) to dust-attenuation-corrected intrinsic UV continua \citep[see e.g.][]{PrietoLyon2023}. Since these emission lines are produced by the excitation of hydrogen or oxygen gas from ionising radiation that does not escape the galaxy, the values inferred from observations correspond actually to $\xi_\mathrm{ion}(1-f_\mathrm{esc})$. Therefore, our predictions for $\xi_\mathrm{ion}$ can be thought of as upper limits for observationally inferred values, though only when the assumed dust attenuation of the UV remains below its actual values. Fig.~\ref{fig_xiion-MUV} shows the relationship between $\xi_\mathrm{ion}$ and observed UV luminosity ($M_\mathrm{UV}$) for the Salpeter IMF (red contours) and the Evolving IMF (blue contours).

We can see that the maximum $\xi_\mathrm{ion}$ value is approximately $10^{25.4}$~Hz~erg$^{-1}$ when assuming a Salpeter IMF. However, when adopting a top-heavy IMF, ranging up to a log-flat IMF as in the Evolving IMF model, $\xi_\mathrm{ion}$ can reach higher values for UV-bright galaxies, up to around $10^{25.7}$~Hz~erg$^{-1}$. The higher abundance of massive stars in a top-heavy IMF shifts stellar emission to higher energies, thereby raising the ratio of \HI ionising photons to UV continuum photons and thus $\xi_\mathrm{ion}$ for highly star-forming, UV-bright galaxies. Compared to the Salpeter IMF, $\xi_\mathrm{ion}$ values in the Evolving IMF model reach about $2\times$ higher values. Here, it is important to note that our $\xi_\mathrm{ion}$ values correspond to results derived under the assumption of continuous star formation over the specified time step ($\sim11$~Myrs at $z=13$). If star formation was stochastic on shorter time scales, $\xi_\mathrm{ion}$ could reach higher values during periods of intense starbursts, presumably up to $10^{26.1}$~Hz~erg$^{-1}$ and $10^{25.8}$~Hz~erg$^{-1}$ for the Evolving and Salpeter IMF, respectively.

Hence, $\xi_\mathrm{ion}$ values exceeding $10^{25.8}$Hz~erg$^{-1}$ are indicative of star formation driven by a top-heavy IMF. For example, JADES-z12-0 \citep{CurtisLake2023} and JADES-z13-1-LA \citep[$\xi_\mathrm{ion}\simeq10^{26.5}$Hz~erg$^{-1}$][]{Witstok2024} — for which our Evolving IMF model predicts a top-heavy IMF based on their SFR and stellar mass values — show $\xi_\mathrm{ion}$ values that are \st{more} consistent with or in case of JADES-z13-1-LA even exceed our Evolving IMF model. Observations at slightly lower redshifts indicate that $\xi_\mathrm{ion}$ tends to increase with redshift \citep{Simmonds2024a, Simmonds2024c}, with some galaxies at $z \simeq 7-9$ surpassing the $\xi_\mathrm{ion}$ threshold for a top-heavy IMF \citep{Endsley2023, Whitler2024, Mascia2024}, though not reaching the extreme value observed for JADES-z13-1-LA at $z=13$. This trend is consistent with our Evolving IMF model, which predicts galaxies with top-heavy IMFs becoming more prevalent at earlier cosmic times.

\section{Conclusions}
\label{sec_conclusions}

We have investigated whether an IMF parameterisation that links the top-heaviness of the IMF to the sSFR of a galaxy, using the gas-to-DM mass ratio as a proxy, can explain the observed abundance of bright $z>10$ galaxies. Specifically, we have considered an IMF that becomes only top-heavy when the sSFR exceeds a certain threshold. To model such a varying IMF, we adopted the IMF shape from \citet{Cueto2024}, using a Salpeter IMF below and a log-flat IMF above a stellar mass $M_c$. Increasing $M_c$ results in a more top-heavy IMF, characterised by a higher fraction of stellar mass produced in massive stars ($M>M_c$), denoted as $f_\mathrm{massive}$.
Using a toy model, we have compared how this Evolving IMF affects galaxy properties, particularly UV luminosities, and their statistical distributions, relative to the standard assumption of a constant Salpeter IMF. Our findings are as follows:
\begin{enumerate}
    \item A more top-heavy IMF reduces the SFE of SN feedback-limited galaxies by the factor by which the energy released by SNe is enhanced. Since the SFE of SN feedback-limited galaxies decreases with the abundance of massive stars, the UV luminosity enhancement for such galaxies does not exceed a factor of $\sim3.5$, corresponding to a shift of $\Delta M_\mathrm{UV} = 1.3$. This enhancement level is nearly achieved with a massive star fraction of $f_\mathrm{massive} \approx 0.5$. Further increasing $f_\mathrm{massive}$ to $1$ results in only an additional $10\%$ enhancement. However, increasing $f_\mathrm{massive}$ shifts the halo mass below which the star formation becomes SN feedback-limited by $\sim0.8$~dex for a change of $\Delta f_\mathrm{massive}=1$.
    \item For more massive galaxies, however, the SFE remains unchanged, leading to a continuous increase in UV luminosity as $f_\mathrm{massive}$ rises, reaching a maximum enhancement of $\sim11$, which corresponds to a shift of $\Delta M_\mathrm{UV} = 2.6$.
    \item Our Evolving IMF model successfully reproduces the observed UV LF at $z=5-15$ due to a growing fraction of galaxies forming stars according to a top-heavy IMF and these IMFs becoming progressively top-heavier towards increasing redshifts. The shallower slopes of the $z>10$ UV LFs in the Evolving IMF model arise from the increasing gas fraction (triggering star formation with top-heavy IMFs) and UV luminosity enhancement with rising halo mass.
    \item In our Evolving IMF model, UV-bright galaxies at $z>10$ fall into two categories: galaxies with the highest stellar and dust masses and galaxies with lower stellar and dust masses whose brightness results from their stars forming according to a top-heavy IMF.
    \item Galaxies forming stars according to a top-heavy IMF are typically the most gas-rich and highly star-forming, with sSFR$\gtrsim25$~Gyr$^{-1}$. These galaxies are located in environments where they form the deepest potential wells or most massive galaxies, allowing for high gas accretion rates replenishing their gas reservoirs and counteracting losses due to SNe.
    \item Adopting our Evolving IMF model increases the $1\sigma$ UV luminosity variance (for a given halo mass) only moderately, not sufficient to reproduce the $z>10$ UV LFs as shown in \citet{Gelli2024, Munoz2023, Kravtsov2024}. However, it is the significant increase (by a factor $1.4-2$) in the $2\sigma$ UV luminosity variance, driven by the UV luminosity enhancement of the most star-forming galaxies, that reproduces the high abundance of UV-bright $z>10$ galaxies.
    \item Adopting an Evolving IMF causes reionisation to start earlier without altering completion time. It also modifies the ionisation morphology during the early stages of reionisation, leading to more extended large ionised regions and fewer small ones.
\end{enumerate}

ogether with \citet{Cueto2024}, our results highlight how the dependence of the IMF on galaxy properties affects the abundance of UV-bright galaxies at $z>10$. Specifically, they suggest that to reproduce the UV LFs at $z=5-15$ with a top-heavy IMF, the IMF needs to be more top-heavy for massive than less-massive galaxies. If the IMF is less top-heavy in massive galaxies, the UV luminosity enhancement becomes similar across all galaxy masses, leading to a UV LF slope at $z>10$ that is steeper than observed.
Increasing the top-heaviness of the IMF with rising sSFR or gas content -- consistent with findings in local observations of star-forming regions and in broad agreement with the IMF trends proposed in the integrated galaxy-wide IMF theory \citep[e.g.][]{Weidner2013, Jerabkova2018} -- results in a larger fraction of more massive galaxies with a top-heavy IMF and reproduces the high-redshift UV LFs without necessitating an increase in the SFE or stochasticity beyond what is accounted for by our standard model.
Interestingly, according to our model, observed galaxies that classify as those with a top-heavy IMF also display signatures of massive stars, such as elevated ionising emissivities and [N/O] abundances. 
Furthermore, adopting a top-heavy IMF in galaxies within overdense regions — where gas accretion rates and SFRs are highest — may help explain the observed clustering bias, which exceeds simple analytic estimates \citep[e.g.][]{Gelli2024}. However, the greater variance in UV luminosity could offset this effect, which we will explore in future work. 
In this context, it is interesting to note that simulations suggest that mergers can trigger higher sSFR values or starbursts \citep[e.g.][]{Croton2016, Patton2020}. Thus, top-heavy IMFs could be predominantly found in merging galaxies, particularly since the galaxy merger rate increases towards higher redshifts \citep[e.g.][]{Gottloeber2001} -- a hypothesis we will test in subsequent work.

Recently, \citet{vanDokkumConroy2024} proposed a concordance IMF for massive galaxies at $z\simeq0-5$, which becomes increasingly both bottom-heavy and top-heavy compared to the Salpeter IMF as the galaxy's velocity dispersion increases. They suggest this IMF could help explain the bright galaxies at $z>7$. This IMF exhibits a similar trend to our Evolving IMF, where more compact (and likely more star-forming) galaxies have more top-heavy IMFs. However, while the shape of the concordance IMF is similar to that of our Evolving IMF in highly star-forming galaxies, its steeper high-mass slope leads to a lower UV boost. Whether this reduced UV boost can be compensated by a higher fraction of galaxies with top-heavy IMFs remains to be tested.

Finally, we note the main limitations of this study.
Firstly, we do not adjust the SFE of massive galaxies with a top-heavy IMF. Simulations of star-forming clouds suggest that a higher radiation pressure on dust grains, caused by an increased abundance of massive stars, can reverse gas accretion and drive gas outward, reducing the SFE by nearly a factor of two at solar metallicity ($Z=Z_\odot$) when UV emissivity is enhanced by an order of magnitude \citep{Menon2024}. However, this reduction in SFE decreases as the gas becomes more metal-poor (also resulting in a lower dust abundance), reaching values similar to those in our simulated galaxies ($Z\simeq0.01 Z_\odot$), and nearly disappears in dense clumps ($\Sigma_\mathrm{cl}\simeq3\times10^4~\msun$~pc$^{-2}$). If the SFE of massive galaxies were decreased, it would likely reduce the UV luminosity enhancement in these galaxies. This effect could potentially be counterbalanced by lowering the sSFR threshold above which our model adopts a top-heavy IMF. We plan to explore this in future work.

Secondly, this study explores a specific IMF parameterisation whose shape is motivated by the star-forming cloud simulations by \citet{Chon2022}. Alternative parameterisations, such as those with an increasingly shallower slope to represent a more top-heavy IMF, could lead to a different ratio of massive to less massive stars, potentially altering the UV luminosity enhancement. Another scenario, which is increasingly supported by observations suggesting the presence of very massive stars (>100~$\msun$) at sub-solar metallicities \citep[e.g.,][]{Martins2023, Mestric2023, Wofford2023, Upadhyaya2024}, is that the IMF extends to masses larger than the 100~$\msun$ assumed here. This extension could also result in a UV luminosity enhancement \citep{Martins2022} and, according to \citet{Katz2024}, leave more distinct signatures in the galaxy spectra, such as redder UV slopes of $\beta>-2$.

Thirdly, the spectra for our stellar populations do not account for the effects of binary interactions on stellar evolution. But massive stars, likely to form in dense regions, are expected to often exist in binaries. Nevertheless, as we explored in \citet{Hutter2021a}, using stellar population synthesis models that include binary interactions, such as {\sc bpass}, hardly changes the UV luminosities of galaxies. They only delay the decrease of the ionising photon output following a starburst and could suppress star formation longer due to the associated radiative feedback -- which would effectively lead to an adjustment of {\sc astraeus}'s SFE and SNe wind coupling parameters.

In summary, our Evolving IMF model appears well-aligned with current high-$z$ galaxy observations and observations of local galaxies and star-forming regions. Thus, a top-heavy IMF in strongly star-forming galaxies offers one possible explanation for the high abundance of UV-bright galaxies at $z>10$. Yet, alternative explanations — such as feedback-free starbursts, bursty star formation, cosmic variance, contributions from black holes, or a combination of these factors, including different IMF scenarios — also remain plausible. An analysis of the UV-luminosity-weighted galaxy clustering bias and gas elemental abundances, such as [N/O] and [C/O], will offer further insights into the applicability of the Evolving IMF model.


\begin{acknowledgements}
    We thank the anonymous referee for their careful reading of the manuscript and their comments, which improved the quality of the paper. AH acknowledges support by the VILLUM FONDEN under grant 37459. The Cosmic Dawn Center (DAWN) is funded by the Danish National Research Foundation under grant DNRF140. 
    PD acknowledges support from NWO grant 016.VIDI.189.162 (``ODIN") and warmly thanks the European Commission and the University of Groningen's CO-FUND Rosalind Franklin program. 
    GY acknowledges  Ministerio de  Ciencia e Innovaci\'on (Spain) for partial financial support under research grant PID2021-122603NB-C21. 
    The {\sc vsmdpl} simulation has been performed at LRZ Munich within the project  pr87yi. The authors gratefully acknowledge the Gauss Centre for Supercomputing e.V. (www.gauss-centre.eu) for funding this project by providing computing time on the GCS Supercomputer SUPERMUC-NG at Leibniz Supercomputing Centre (www.lrz.de). The CosmoSim database (\url{www.cosmosim.org}) provides access to the simulation and the Rockstar data. The database is a service by the Leibniz Institute for Astrophysics Potsdam (AIP). 
    This research made use of \texttt{matplotlib}, a Python library for publication-quality graphics \citep{hunter2007}; and the Python library \texttt{numpy} \citep{numpy}.
\end{acknowledgements}

\bibliographystyle{aa} 
\bibliography{ref}

\begin{thebibliography}{151}
\expandafter\ifx\csname natexlab\endcsname\relax\def\natexlab#1{#1}\fi

\bibitem[{{Abel} {et~al.}(2002){Abel}, {Bryan}, \& {Norman}}]{Abel2002}
{Abel}, T., {Bryan}, G.~L., \& {Norman}, M.~L. 2002, Science, 295, 93

\bibitem[{{Adams} {et~al.}(2024){Adams}, {Conselice}, {Austin}, {Harvey},
  {Ferreira}, {Trussler}, {Juod{\v{z}}balis}, {Li}, {Windhorst}, {Cohen},
  {Jansen}, {Summers}, {Tompkins}, {Driver}, {Robotham}, {D'Silva}, {Yan},
  {Coe}, {Frye}, {Grogin}, {Koekemoer}, {Marshall}, {Pirzkal}, {Ryan},
  {Maksym}, {Rutkowski}, {Willmer}, {Hammel}, {Nonino}, {Bhatawdekar},
  {Wilkins}, {Bradley}, {Broadhurst}, {Cheng}, {Dole}, {Hathi}, \&
  {Zitrin}}]{Adams2024}
{Adams}, N.~J., {Conselice}, C.~J., {Austin}, D., {et~al.} 2024, \apj, 965, 169

\bibitem[{{Adams} {et~al.}(2023){Adams}, {Conselice}, {Ferreira}, {Austin},
  {Trussler}, {Juod{\v{z}}balis}, {Wilkins}, {Caruana}, {Dayal}, {Verma}, \&
  {Vijayan}}]{Adams2023}
{Adams}, N.~J., {Conselice}, C.~J., {Ferreira}, L., {et~al.} 2023, \mnras, 518,
  4755

\bibitem[{{Arrabal Haro} {et~al.}(2023){Arrabal Haro}, {Dickinson},
  {Finkelstein}, {Kartaltepe}, {Donnan}, {Burgarella}, {Carnall}, {Cullen},
  {Dunlop}, {Fern{\'a}ndez}, {Fujimoto}, {Jung}, {Krips}, {Larson}, {Papovich},
  {P{\'e}rez-Gonz{\'a}lez}, {Amor{\'\i}n}, {Bagley}, {Buat}, {Casey},
  {Chworowsky}, {Cohen}, {Ferguson}, {Giavalisco}, {Huertas-Company},
  {Hutchison}, {Kocevski}, {Koekemoer}, {Lucas}, {McLeod}, {McLure}, {Pirzkal},
  {Seill{\'e}}, {Trump}, {Weiner}, {Wilkins}, \& {Zavala}}]{ArrabalHaro2023}
{Arrabal Haro}, P., {Dickinson}, M., {Finkelstein}, S.~L., {et~al.} 2023, \nat,
  622, 707

\bibitem[{{Atek} {et~al.}(2023){Atek}, {Chemerynska}, {Wang}, {Furtak},
  {Weibel}, {Oesch}, {Weaver}, {Labb{\'e}}, {Bezanson}, {van Dokkum}, {Zitrin},
  {Dayal}, {Williams}, {Nannayakkara}, {Price}, {Brammer}, {Goulding}, {Leja},
  {Marchesini}, {Nelson}, {Pan}, \& {Whitaker}}]{Atek2023}
{Atek}, H., {Chemerynska}, I., {Wang}, B., {et~al.} 2023, \mnras, 524, 5486

\bibitem[{{Atek} {et~al.}(2024){Atek}, {Labb{\'e}}, {Furtak}, {Chemerynska},
  {Fujimoto}, {Setton}, {Miller}, {Oesch}, {Bezanson}, {Price}, {Dayal},
  {Zitrin}, {Kokorev}, {Weaver}, {Brammer}, {Dokkum}, {Williams}, {Cutler},
  {Feldmann}, {Fudamoto}, {Greene}, {Leja}, {Maseda}, {Muzzin}, {Pan},
  {Papovich}, {Nelson}, {Nanayakkara}, {Stark}, {Stefanon}, {Suess}, {Wang}, \&
  {Whitaker}}]{Atek2024}
{Atek}, H., {Labb{\'e}}, I., {Furtak}, L.~J., {et~al.} 2024, \nat, 626, 975

\bibitem[{{Atek} {et~al.}(2015){Atek}, {Richard}, {Kneib}, {Jauzac},
  {Schaerer}, {Clement}, {Limousin}, {Jullo}, {Natarajan}, {Egami}, \&
  {Ebeling}}]{Atek2015}
{Atek}, H., {Richard}, J., {Kneib}, J.-P., {et~al.} 2015, \apj, 800, 18

\bibitem[{{Atek} {et~al.}(2018){Atek}, {Richard}, {Kneib}, \&
  {Schaerer}}]{Atek2018}
{Atek}, H., {Richard}, J., {Kneib}, J.-P., \& {Schaerer}, D. 2018, \mnras, 479,
  5184

\bibitem[{{Austin} {et~al.}(2023){Austin}, {Adams}, {Conselice}, {Harvey},
  {Ormerod}, {Trussler}, {Li}, {Ferreira}, {Dayal}, \&
  {Juod{\v{z}}balis}}]{Austin2023}
{Austin}, D., {Adams}, N., {Conselice}, C.~J., {et~al.} 2023, \apjl, 952, L7

\bibitem[{{Bakx} {et~al.}(2023){Bakx}, {Zavala}, {Mitsuhashi}, {Treu},
  {Fontana}, {Tadaki}, {Casey}, {Castellano}, {Glazebrook}, {Hagimoto},
  {Ikeda}, {Jones}, {Leethochawalit}, {Mason}, {Morishita}, {Nanayakkara},
  {Pentericci}, {Roberts-Borsani}, {Santini}, {Serjeant}, {Tamura}, {Trenti},
  \& {Vanzella}}]{Bakx2023}
{Bakx}, T. J.~L.~C., {Zavala}, J.~A., {Mitsuhashi}, I., {et~al.} 2023, \mnras,
  519, 5076

\bibitem[{{Barkana} \& {Loeb}(2001)}]{Barkana2001}
{Barkana}, R. \& {Loeb}, A. 2001, \physrep, 349, 125

\bibitem[{{Bolan} {et~al.}(2022){Bolan}, {Lemaux}, {Mason}, {Brada{\v{c}}},
  {Treu}, {Strait}, {Pelliccia}, {Pentericci}, \& {Malkan}}]{Bolan2022}
{Bolan}, P., {Lemaux}, B.~C., {Mason}, C., {et~al.} 2022, \mnras, 517, 3263

\bibitem[{{Bosman} {et~al.}(2022){Bosman}, {Davies}, {Becker}, {Keating},
  {Davies}, {Zhu}, {Eilers}, {D'Odorico}, {Bian}, {Bischetti}, {Cristiani},
  {Fan}, {Farina}, {Haehnelt}, {Hennawi}, {Kulkarni}, {Mesinger}, {Meyer},
  {Onoue}, {Pallottini}, {Qin}, {Ryan-Weber}, {Schindler}, {Walter}, {Wang}, \&
  {Yang}}]{Bosman2022}
{Bosman}, S. E.~I., {Davies}, F.~B., {Becker}, G.~D., {et~al.} 2022, \mnras,
  514, 55

\bibitem[{{Bouwens} {et~al.}(2023{\natexlab{a}}){Bouwens}, {Illingworth},
  {Oesch}, {Stefanon}, {Naidu}, {van Leeuwen}, \& {Magee}}]{Bouwens2023a}
{Bouwens}, R., {Illingworth}, G., {Oesch}, P., {et~al.} 2023{\natexlab{a}},
  \mnras, 523, 1009

\bibitem[{{Bouwens} {et~al.}(2021){Bouwens}, {Oesch}, {Stefanon},
  {Illingworth}, {Labb{\'e}}, {Reddy}, {Atek}, {Montes}, {Naidu},
  {Nanayakkara}, {Nelson}, \& {Wilkins}}]{Bouwens2021}
{Bouwens}, R.~J., {Oesch}, P.~A., {Stefanon}, M., {et~al.} 2021, \aj, 162, 47

\bibitem[{{Bouwens} {et~al.}(2023{\natexlab{b}}){Bouwens}, {Stefanon},
  {Brammer}, {Oesch}, {Herard-Demanche}, {Illingworth}, {Matthee}, {Naidu},
  {van Dokkum}, \& {van Leeuwen}}]{Bouwens2023b}
{Bouwens}, R.~J., {Stefanon}, M., {Brammer}, G., {et~al.} 2023{\natexlab{b}},
  \mnras, 523, 1036

\bibitem[{{Boylan-Kolchin}(2023)}]{BoylanKolchin2023}
{Boylan-Kolchin}, M. 2023, Nature Astronomy, 7, 731

\bibitem[{{Bradley} {et~al.}(2023){Bradley}, {Coe}, {Brammer}, {Furtak},
  {Larson}, {Kokorev}, {Andrade-Santos}, {Bhatawdekar}, {Brada{\v{c}}},
  {Broadhurst}, {Carnall}, {Conselice}, {Diego}, {Frye}, {Fujimoto}, {Hsiao},
  {Hutchison}, {Jung}, {Mahler}, {McCandliss}, {Oguri}, {Postman}, {Sharon},
  {Trenti}, {Vanzella}, {Welch}, {Windhorst}, \& {Zitrin}}]{Bradley2023}
{Bradley}, L.~D., {Coe}, D., {Brammer}, G., {et~al.} 2023, \apj, 955, 13

\bibitem[{{Bromm} {et~al.}(2002){Bromm}, {Coppi}, \& {Larson}}]{Bromm2002}
{Bromm}, V., {Coppi}, P.~S., \& {Larson}, R.~B. 2002, \apj, 564, 23

\bibitem[{{Bunker} {et~al.}(2023){Bunker}, {Saxena}, {Cameron}, {Willott},
  {Curtis-Lake}, {Jakobsen}, {Carniani}, {Smit}, {Maiolino}, {Witstok},
  {Curti}, {D'Eugenio}, {Jones}, {Ferruit}, {Arribas}, {Charlot}, {Chevallard},
  {Giardino}, {de Graaff}, {Looser}, {L{\"u}tzgendorf}, {Maseda}, {Rawle},
  {Rix}, {Del Pino}, {Alberts}, {Egami}, {Eisenstein}, {Endsley}, {Hainline},
  {Hausen}, {Johnson}, {Rieke}, {Rieke}, {Robertson}, {Shivaei}, {Stark},
  {Sun}, {Tacchella}, {Tang}, {Williams}, {Willmer}, {Baker}, {Baum},
  {Bhatawdekar}, {Bowler}, {Boyett}, {Chen}, {Circosta}, {Helton}, {Ji},
  {Kumari}, {Lyu}, {Nelson}, {Parlanti}, {Perna}, {Sandles}, {Scholtz},
  {Suess}, {Topping}, {{\"U}bler}, {Wallace}, \& {Whitler}}]{Bunker2023}
{Bunker}, A.~J., {Saxena}, A., {Cameron}, A.~J., {et~al.} 2023, \aap, 677, A88

\bibitem[{{Cameron} {et~al.}(2024){Cameron}, {Katz}, {Witten}, {Saxena},
  {Laporte}, \& {Bunker}}]{Cameron2024}
{Cameron}, A.~J., {Katz}, H., {Witten}, C., {et~al.} 2024, \mnras, 534, 523

\bibitem[{{Caruana} {et~al.}(2012){Caruana}, {Bunker}, {Wilkins}, {Stanway},
  {Lacy}, {Jarvis}, {Lorenzoni}, \& {Hickey}}]{Caruana2012}
{Caruana}, J., {Bunker}, A.~J., {Wilkins}, S.~M., {et~al.} 2012, \mnras, 427,
  3055

\bibitem[{{Caruana} {et~al.}(2014){Caruana}, {Bunker}, {Wilkins}, {Stanway},
  {Lorenzoni}, {Jarvis}, \& {Ebert}}]{Caruana2014}
{Caruana}, J., {Bunker}, A.~J., {Wilkins}, S.~M., {et~al.} 2014, \mnras, 443,
  2831

\bibitem[{{Castellano} {et~al.}(2010){Castellano}, {Fontana}, {Paris},
  {Grazian}, {Pentericci}, {Boutsia}, {Santini}, {Testa}, {Dickinson},
  {Giavalisco}, {Bouwens}, {Cuby}, {Mannucci}, {Cl{\'e}ment}, {Cristiani},
  {Fiore}, {Gallozzi}, {Giallongo}, {Maiolino}, {Menci}, {Moorwood}, {Nonino},
  {Renzini}, {Rosati}, {Salimbeni}, \& {Vanzella}}]{Castellano2010}
{Castellano}, M., {Fontana}, A., {Paris}, D., {et~al.} 2010, \aap, 524, A28

\bibitem[{{Charbonnel} {et~al.}(2023){Charbonnel}, {Schaerer}, {Prantzos},
  {Ram{\'\i}rez-Galeano}, {Fragos}, {Kuruvanthodi}, {Marques-Chaves}, \&
  {Gieles}}]{Charbonnel2023}
{Charbonnel}, C., {Schaerer}, D., {Prantzos}, N., {et~al.} 2023, \aap, 673, L7

\bibitem[{{Chon} {et~al.}(2022){Chon}, {Ono}, {Omukai}, \&
  {Schneider}}]{Chon2022}
{Chon}, S., {Ono}, H., {Omukai}, K., \& {Schneider}, R. 2022, \mnras, 514, 4639

\bibitem[{{Clark} {et~al.}(2011){Clark}, {Glover}, {Klessen}, \&
  {Bromm}}]{Clark2011}
{Clark}, P.~C., {Glover}, S. C.~O., {Klessen}, R.~S., \& {Bromm}, V. 2011,
  \apj, 727, 110

\bibitem[{{Croton} {et~al.}(2016){Croton}, {Stevens}, {Tonini}, {Garel},
  {Bernyk}, {Bibiano}, {Hodkinson}, {Mutch}, {Poole}, \&
  {Shattow}}]{Croton2016}
{Croton}, D.~J., {Stevens}, A. R.~H., {Tonini}, C., {et~al.} 2016, \apjs, 222,
  22

\bibitem[{{Cueto} {et~al.}(2024){Cueto}, {Hutter}, {Dayal}, {Gottl{\"o}ber},
  {Heintz}, {Mason}, {Trebitsch}, \& {Yepes}}]{Cueto2024}
{Cueto}, E.~R., {Hutter}, A., {Dayal}, P., {et~al.} 2024, \aap, 686, A138

\bibitem[{{Cullen} {et~al.}(2024){Cullen}, {McLeod}, {McLure}, {Dunlop},
  {Donnan}, {Carnall}, {Keating}, {Magee}, {Arellano-Cordova}, {Bowler},
  {Begley}, {Flury}, {Hamadouche}, \& {Stanton}}]{Cullen2024}
{Cullen}, F., {McLeod}, D.~J., {McLure}, R.~J., {et~al.} 2024, \mnras, 531, 997

\bibitem[{{Curti} {et~al.}(2024){Curti}, {Witstok}, {Jakobsen}, {Kobayashi},
  {Curtis-Lake}, {Hainline}, {Ji}, {D'Eugenio}, {Chevallard}, {Maiolino},
  {Scholtz}, {Carniani}, {Arribas}, {Baker}, {Bhatawdekar}, {Boyett}, {Bunker},
  {Cameron}, {Cargile}, {Charlot}, {Eisenstein}, {Ji}, {Johnson}, {Kumari},
  {Maseda}, {Robertson}, {Silcock}, {Tacchella}, {Ubler}, {Venturi},
  {Williams}, {Willmer}, \& {Willott}}]{Curti2024}
{Curti}, M., {Witstok}, J., {Jakobsen}, P., {et~al.} 2024, arXiv e-prints,
  arXiv:2407.02575

\bibitem[{{Curtis-Lake} {et~al.}(2023){Curtis-Lake}, {Carniani}, {Cameron},
  {Charlot}, {Jakobsen}, {Maiolino}, {Bunker}, {Witstok}, {Smit}, {Chevallard},
  {Willott}, {Ferruit}, {Arribas}, {Bonaventura}, {Curti}, {D'Eugenio},
  {Franx}, {Giardino}, {Looser}, {L{\"u}tzgendorf}, {Maseda}, {Rawle}, {Rix},
  {Rodr{\'\i}guez del Pino}, {{\"U}bler}, {Sirianni}, {Dressler}, {Egami},
  {Eisenstein}, {Endsley}, {Hainline}, {Hausen}, {Johnson}, {Rieke},
  {Robertson}, {Shivaei}, {Stark}, {Tacchella}, {Williams}, {Willmer},
  {Bhatawdekar}, {Bowler}, {Boyett}, {Chen}, {de Graaff}, {Helton}, {Hviding},
  {Jones}, {Kumari}, {Lyu}, {Nelson}, {Perna}, {Sandles}, {Saxena}, {Suess},
  {Sun}, {Topping}, {Wallace}, \& {Whitler}}]{CurtisLake2023}
{Curtis-Lake}, E., {Carniani}, S., {Cameron}, A., {et~al.} 2023, Nature
  Astronomy, 7, 622

\bibitem[{{Davies} {et~al.}(2018){Davies}, {Hennawi}, {Ba{\~n}ados},
  {Luki{\'c}}, {Decarli}, {Fan}, {Farina}, {Mazzucchelli}, {Rix}, {Venemans},
  {Walter}, {Wang}, \& {Yang}}]{Davies2018}
{Davies}, F.~B., {Hennawi}, J.~F., {Ba{\~n}ados}, E., {et~al.} 2018, \apj, 864,
  142

\bibitem[{{Dayal} \& {Ferrara}(2018)}]{Dayal2018}
{Dayal}, P. \& {Ferrara}, A. 2018, \physrep, 780, 1

\bibitem[{{Dayal} {et~al.}(2014){Dayal}, {Ferrara}, {Dunlop}, \&
  {Pacucci}}]{Dayal2014}
{Dayal}, P., {Ferrara}, A., {Dunlop}, J.~S., \& {Pacucci}, F. 2014, \mnras,
  445, 2545

\bibitem[{{Dayal} {et~al.}(2022){Dayal}, {Ferrara}, {Sommovigo}, {Bouwens},
  {Oesch}, {Smit}, {Gonzalez}, {Schouws}, {Stefanon}, {Kobayashi}, {Bremer},
  {Algera}, {Aravena}, {Bowler}, {da Cunha}, {Fudamoto}, {Graziani}, {Hodge},
  {Inami}, {De Looze}, {Pallottini}, {Riechers}, {Schneider}, {Stark}, \&
  {Endsley}}]{Dayal2022}
{Dayal}, P., {Ferrara}, A., {Sommovigo}, L., {et~al.} 2022, \mnras, 512, 989

\bibitem[{{Dekel} {et~al.}(2023){Dekel}, {Sarkar}, {Birnboim}, {Mandelker}, \&
  {Li}}]{Dekel2023}
{Dekel}, A., {Sarkar}, K.~C., {Birnboim}, Y., {Mandelker}, N., \& {Li}, Z.
  2023, \mnras, 523, 3201

\bibitem[{{Donnan} {et~al.}(2023{\natexlab{a}}){Donnan}, {McLeod}, {Dunlop},
  {McLure}, {Carnall}, {Begley}, {Cullen}, {Hamadouche}, {Bowler}, {Magee},
  {McCracken}, {Milvang-Jensen}, {Moneti}, \& {Targett}}]{Donnan2023a}
{Donnan}, C.~T., {McLeod}, D.~J., {Dunlop}, J.~S., {et~al.} 2023{\natexlab{a}},
  \mnras, 518, 6011

\bibitem[{{Donnan} {et~al.}(2023{\natexlab{b}}){Donnan}, {McLeod}, {McLure},
  {Dunlop}, {Carnall}, {Cullen}, \& {Magee}}]{Donnan2023b}
{Donnan}, C.~T., {McLeod}, D.~J., {McLure}, R.~J., {et~al.} 2023{\natexlab{b}},
  \mnras, 520, 4554

\bibitem[{{Donnan} {et~al.}(2024){Donnan}, {McLure}, {Dunlop}, {McLeod},
  {Magee}, {Arellano-C{\'o}rdova}, {Barrufet}, {Begley}, {Bowler}, {Carnall},
  {Cullen}, {Ellis}, {Fontana}, {Illingworth}, {Grogin}, {Hamadouche},
  {Koekemoer}, {Liu}, {Mason}, {Santini}, \& {Stanton}}]{Donnan2024}
{Donnan}, C.~T., {McLure}, R.~J., {Dunlop}, J.~S., {et~al.} 2024, \mnras, 533,
  3222

\bibitem[{{Duncan} {et~al.}(2014){Duncan}, {Conselice}, {Mortlock}, {Hartley},
  {Guo}, {Ferguson}, {Dav{\'e}}, {Lu}, {Ownsworth}, {Ashby}, {Dekel},
  {Dickinson}, {Faber}, {Giavalisco}, {Grogin}, {Kocevski}, {Koekemoer},
  {Somerville}, \& {White}}]{duncan_mass_2014}
{Duncan}, K., {Conselice}, C.~J., {Mortlock}, A., {et~al.} 2014, \mnras, 444,
  2960

\bibitem[{{Ekstr{\"o}m}(2021)}]{Ekstrom2021}
{Ekstr{\"o}m}, S. 2021, Frontiers in Astronomy and Space Sciences, 8, 53

\bibitem[{{Endsley} {et~al.}(2023){Endsley}, {Stark}, {Whitler}, {Topping},
  {Chen}, {Plat}, {Chisholm}, \& {Charlot}}]{Endsley2023}
{Endsley}, R., {Stark}, D.~P., {Whitler}, L., {et~al.} 2023, \mnras, 524, 2312

\bibitem[{{Fan} {et~al.}(2006){Fan}, {Strauss}, {Becker}, {White}, {Gunn},
  {Knapp}, {Richards}, {Schneider}, {Brinkmann}, \& {Fukugita}}]{Fan2006}
{Fan}, X., {Strauss}, M.~A., {Becker}, R.~H., {et~al.} 2006, \aj, 132, 117

\bibitem[{{Ferrara} {et~al.}(2023){Ferrara}, {Pallottini}, \&
  {Dayal}}]{Ferrara2023}
{Ferrara}, A., {Pallottini}, A., \& {Dayal}, P. 2023, \mnras, 522, 3986

\bibitem[{{Finkelstein} {et~al.}(2023){Finkelstein}, {Bagley}, {Ferguson},
  {Wilkins}, {Kartaltepe}, {Papovich}, {Yung}, {Arrabal Haro}, {Behroozi},
  {Dickinson}, {Kocevski}, {Koekemoer}, {Larson}, {Le Bail}, {Morales},
  {P{\'e}rez-Gonz{\'a}lez}, {Burgarella}, {Dav{\'e}}, {Hirschmann},
  {Somerville}, {Wuyts}, {Bromm}, {Casey}, {Fontana}, {Fujimoto}, {Gardner},
  {Giavalisco}, {Grazian}, {Grogin}, {Hathi}, {Hutchison}, {Jha}, {Jogee},
  {Kewley}, {Kirkpatrick}, {Long}, {Lotz}, {Pentericci}, {Pierel}, {Pirzkal},
  {Ravindranath}, {Ryan}, {Trump}, {Yang}, {Bhatawdekar}, {Bisigello}, {Buat},
  {Calabr{\`o}}, {Castellano}, {Cleri}, {Cooper}, {Croton}, {Daddi}, {Dekel},
  {Elbaz}, {Franco}, {Gawiser}, {Holwerda}, {Huertas-Company}, {Jaskot},
  {Leung}, {Lucas}, {Mobasher}, {Pandya}, {Tacchella}, {Weiner}, \&
  {Zavala}}]{Finkelstein2023}
{Finkelstein}, S.~L., {Bagley}, M.~B., {Ferguson}, H.~C., {et~al.} 2023, \apjl,
  946, L13

\bibitem[{{Fiore} {et~al.}(2023){Fiore}, {Ferrara}, {Bischetti}, {Feruglio}, \&
  {Travascio}}]{Fiore2023}
{Fiore}, F., {Ferrara}, A., {Bischetti}, M., {Feruglio}, C., \& {Travascio}, A.
  2023, \apjl, 943, L27

\bibitem[{{Fontanot} {et~al.}(2017){Fontanot}, {De Lucia}, {Hirschmann},
  {Bruzual}, {Charlot}, \& {Zibetti}}]{Fontanot2017}
{Fontanot}, F., {De Lucia}, G., {Hirschmann}, M., {et~al.} 2017, \mnras, 464,
  3812

\bibitem[{{Fontanot} {et~al.}(2018){Fontanot}, {De Lucia}, {Xie}, {Hirschmann},
  {Bruzual}, \& {Charlot}}]{Fontanot2018a}
{Fontanot}, F., {De Lucia}, G., {Xie}, L., {et~al.} 2018, \mnras, 475, 2467

\bibitem[{{Fudamoto} {et~al.}(2022){Fudamoto}, {Smit}, {Bowler}, {Oesch},
  {Bouwens}, {Stefanon}, {Inami}, {Endsley}, {Gonzalez}, {Schouws}, {Stark},
  {Algera}, {Aravena}, {Barrufet}, {da Cunha}, {Dayal}, {Ferrara}, {Graziani},
  {Hodge}, {Hygate}, {Inoue}, {Nanayakkara}, {Pallottini}, {Pizzati},
  {Schneider}, {Sommovigo}, {Sugahara}, {Topping}, {van der Werf}, {Bethermin},
  {Cassata}, {Dessauges-Zavadsky}, {Ibar}, {Faisst}, {Fujimoto}, {Ginolfi},
  {Hathi}, {Jones}, {Pozzi}, \& {Schaerer}}]{Fudamoto2022}
{Fudamoto}, Y., {Smit}, R., {Bowler}, R.~A.~A., {et~al.} 2022, \apj, 934, 144

\bibitem[{{Fujimoto} {et~al.}(2020){Fujimoto}, {Silverman}, {Bethermin},
  {Ginolfi}, {Jones}, {Le F{\`e}vre}, {Dessauges-Zavadsky}, {Rujopakarn},
  {Faisst}, {Fudamoto}, {Cassata}, {Morselli}, {Maiolino}, {Schaerer}, {Capak},
  {Yan}, {Vallini}, {Toft}, {Loiacono}, {Zamorani}, {Talia}, {Narayanan},
  {Hathi}, {Lemaux}, {Boquien}, {Amorin}, {Ibar}, {Koekemoer},
  {M{\'e}ndez-Hern{\'a}ndez}, {Bardelli}, {Vergani}, {Zucca}, {Romano}, \&
  {Cimatti}}]{Fujimoto2020}
{Fujimoto}, S., {Silverman}, J.~D., {Bethermin}, M., {et~al.} 2020, \apj, 900,
  1

\bibitem[{{Fukushima} {et~al.}(2020){Fukushima}, {Hosokawa}, {Chiaki},
  {Omukai}, {Yoshida}, \& {Kuiper}}]{Fukushima2020}
{Fukushima}, H., {Hosokawa}, T., {Chiaki}, G., {et~al.} 2020, \mnras, 497, 829

\bibitem[{{Gelli} {et~al.}(2024){Gelli}, {Mason}, \& {Hayward}}]{Gelli2024}
{Gelli}, V., {Mason}, C., \& {Hayward}, C.~C. 2024, \apj, 975, 192

\bibitem[{{Gonz{\'a}lez} {et~al.}(2011){Gonz{\'a}lez}, {Labb{\'e}}, {Bouwens},
  {Illingworth}, {Franx}, \& {Kriek}}]{gonzalez_evolution_2011}
{Gonz{\'a}lez}, V., {Labb{\'e}}, I., {Bouwens}, R.~J., {et~al.} 2011, \apjl,
  735, L34

\bibitem[{{Gottl{\"o}ber} {et~al.}(2001){Gottl{\"o}ber}, {Klypin}, \&
  {Kravtsov}}]{Gottloeber2001}
{Gottl{\"o}ber}, S., {Klypin}, A., \& {Kravtsov}, A.~V. 2001, \apj, 546, 223

\bibitem[{{Greig} {et~al.}(2019){Greig}, {Mesinger}, \&
  {Ba{\~n}ados}}]{Greig2019}
{Greig}, B., {Mesinger}, A., \& {Ba{\~n}ados}, E. 2019, \mnras, 484, 5094

\bibitem[{{Gunawardhana} {et~al.}(2011){Gunawardhana}, {Hopkins}, {Sharp},
  {Brough}, {Taylor}, {Bland-Hawthorn}, {Maraston}, {Tuffs}, {Popescu},
  {Wijesinghe}, {Jones}, {Croom}, {Sadler}, {Wilkins}, {Driver}, {Liske},
  {Norberg}, {Baldry}, {Bamford}, {Loveday}, {Peacock}, {Robotham}, {Zucker},
  {Parker}, {Conselice}, {Cameron}, {Frenk}, {Hill}, {Kelvin}, {Kuijken},
  {Madore}, {Nichol}, {Parkinson}, {Pimbblet}, {Prescott}, {Sutherland},
  {Thomas}, \& {van Kampen}}]{Gunawardhana2011}
{Gunawardhana}, M.~L.~P., {Hopkins}, A.~M., {Sharp}, R.~G., {et~al.} 2011,
  \mnras, 415, 1647

\bibitem[{{Harikane} {et~al.}(2024){Harikane}, {Inoue}, {Ellis}, {Ouchi},
  {Nakazato}, {Yoshida}, {Ono}, {Sun}, {Sato}, {Fujimoto}, {Kashikawa},
  {McLeod}, {Perez-Gonzalez}, {Sawicki}, {Sugahara}, {Xu}, {Yamanaka},
  {Carnall}, {Cullen}, {Dunlop}, {Egami}, {Grogin}, {Isobe}, {Koekemoer},
  {Laporte}, {Lee}, {Magee}, {Matsuo}, {Matsuoka}, {Mawatari}, {Nakajima},
  {Nakane}, {Tamura}, {Umeda}, \& {Yanagisawa}}]{Harikane2024}
{Harikane}, Y., {Inoue}, A.~K., {Ellis}, R.~S., {et~al.} 2024, arXiv e-prints,
  arXiv:2406.18352

\bibitem[{{Haslbauer} {et~al.}(2022){Haslbauer}, {Kroupa}, {Zonoozi}, \&
  {Haghi}}]{Haslbauer2022}
{Haslbauer}, M., {Kroupa}, P., {Zonoozi}, A.~H., \& {Haghi}, H. 2022, \apjl,
  939, L31

\bibitem[{{Heintz} {et~al.}(2023){Heintz}, {Gim{\'e}nez-Arteaga}, {Fujimoto},
  {Brammer}, {Espada}, {Gillman}, {Gonz{\'a}lez-L{\'o}pez}, {Greve},
  {Harikane}, {Hatsukade}, {Knudsen}, {Koekemoer}, {Kohno}, {Kokorev}, {Lee},
  {Magdis}, {Nelson}, {Rizzo}, {Sanders}, {Schaerer}, {Shapley}, {Strait},
  {Toft}, {Valentino}, {van der Wel}, {Vijayan}, {Watson}, {Bauer},
  {Christiansen}, \& {Wilson}}]{Heintz2023a}
{Heintz}, K.~E., {Gim{\'e}nez-Arteaga}, C., {Fujimoto}, S., {et~al.} 2023,
  \apjl, 944, L30

\bibitem[{Hunter(2007)}]{hunter2007}
Hunter, J.~D. 2007, Computing In Science \& Engineering, 9, 90

\bibitem[{{Hutter}(2018)}]{Hutter2018a}
{Hutter}, A. 2018, \mnras, 477, 1549

\bibitem[{{Hutter} {et~al.}(2021){Hutter}, {Dayal}, {Yepes}, {Gottl{\"o}ber},
  {Legrand}, \& {Ucci}}]{Hutter2021a}
{Hutter}, A., {Dayal}, P., {Yepes}, G., {et~al.} 2021, \mnras, 503, 3698

\bibitem[{{Hutter} {et~al.}(2023){Hutter}, {Trebitsch}, {Dayal},
  {Gottl{\"o}ber}, {Yepes}, \& {Legrand}}]{Hutter2023a}
{Hutter}, A., {Trebitsch}, M., {Dayal}, P., {et~al.} 2023, \mnras, 524, 6124

\bibitem[{{Ishigaki} {et~al.}(2018){Ishigaki}, {Kawamata}, {Ouchi}, {Oguri},
  {Shimasaku}, \& {Ono}}]{Ishigaki2018}
{Ishigaki}, M., {Kawamata}, R., {Ouchi}, M., {et~al.} 2018, \apj, 854, 73

\bibitem[{{Isobe} {et~al.}(2023){Isobe}, {Ouchi}, {Tominaga}, {Watanabe},
  {Nakajima}, {Umeda}, {Yajima}, {Harikane}, {Fukushima}, {Xu}, {Ono}, \&
  {Zhang}}]{Isobe2023}
{Isobe}, Y., {Ouchi}, M., {Tominaga}, N., {et~al.} 2023, \apj, 959, 100

\bibitem[{{Je{\v{r}}{\'a}bkov{\'a}} {et~al.}(2018){Je{\v{r}}{\'a}bkov{\'a}},
  {Hasani Zonoozi}, {Kroupa}, {Beccari}, {Yan}, {Vazdekis}, \&
  {Zhang}}]{Jerabkova2018}
{Je{\v{r}}{\'a}bkov{\'a}}, T., {Hasani Zonoozi}, A., {Kroupa}, P., {et~al.}
  2018, \aap, 620, A39

\bibitem[{{Kashikawa} {et~al.}(2011){Kashikawa}, {Shimasaku}, {Matsuda},
  {Egami}, {Jiang}, {Nagao}, {Ouchi}, {Malkan}, {Hattori}, {Ota}, {Taniguchi},
  {Okamura}, {Ly}, {Iye}, {Furusawa}, {Shioya}, {Shibuya}, {Ishizaki}, \&
  {Toshikawa}}]{Kashikawa2011}
{Kashikawa}, N., {Shimasaku}, K., {Matsuda}, Y., {et~al.} 2011, \apj, 734, 119

\bibitem[{{Katz} {et~al.}(2024){Katz}, {Cameron}, {Saxena}, {Barrufet},
  {Choustikov}, {Cleri}, {de Graaff}, {Ellis}, {Fosbury}, {Heintz}, {Maseda},
  {Matthee}, {McConchie}, \& {Oesch}}]{Katz2024}
{Katz}, H., {Cameron}, A.~J., {Saxena}, A., {et~al.} 2024, arXiv e-prints,
  arXiv:2408.03189

\bibitem[{{Kobayashi} \& {Ferrara}(2024)}]{Kobayashi2024}
{Kobayashi}, C. \& {Ferrara}, A. 2024, \apjl, 962, L6

\bibitem[{{Kobayashi} {et~al.}(2020){Kobayashi}, {Karakas}, \&
  {Lugaro}}]{Kobayashi2020b}
{Kobayashi}, C., {Karakas}, A.~I., \& {Lugaro}, M. 2020, \apj, 900, 179

\bibitem[{{Konno} {et~al.}(2018){Konno}, {Ouchi}, {Shibuya}, {Ono},
  {Shimasaku}, {Taniguchi}, {Nagao}, {Kobayashi}, {Kajisawa}, {Kashikawa},
  {Inoue}, {Oguri}, {Furusawa}, {Goto}, {Harikane}, {Higuchi}, {Komiyama},
  {Kusakabe}, {Miyazaki}, {Nakajima}, \& {Wang}}]{Konno2018}
{Konno}, A., {Ouchi}, M., {Shibuya}, T., {et~al.} 2018, \pasj, 70, S16

\bibitem[{{Kravtsov} \& {Belokurov}(2024)}]{Kravtsov2024}
{Kravtsov}, A. \& {Belokurov}, V. 2024, arXiv e-prints, arXiv:2405.04578

\bibitem[{{Kroupa} \& {Weidner}(2003)}]{Kroupa2003}
{Kroupa}, P. \& {Weidner}, C. 2003, \apj, 598, 1076

\bibitem[{{Labb{\'e}} {et~al.}(2023){Labb{\'e}}, {van Dokkum}, {Nelson},
  {Bezanson}, {Suess}, {Leja}, {Brammer}, {Whitaker}, {Mathews}, {Stefanon}, \&
  {Wang}}]{Labbe2023}
{Labb{\'e}}, I., {van Dokkum}, P., {Nelson}, E., {et~al.} 2023, \nat, 616, 266

\bibitem[{{Legrand} {et~al.}(2023){Legrand}, {Dayal}, {Hutter},
  {Gottl{\"o}ber}, {Yepes}, \& {Trebitsch}}]{Legrand2023}
{Legrand}, L., {Dayal}, P., {Hutter}, A., {et~al.} 2023, \mnras, 519, 4564

\bibitem[{{Legrand} {et~al.}(2022){Legrand}, {Hutter}, {Dayal}, {Ucci},
  {Gottl{\"o}ber}, \& {Yepes}}]{Legrand2022}
{Legrand}, L., {Hutter}, A., {Dayal}, P., {et~al.} 2022, \mnras, 509, 595

\bibitem[{{Leitherer} {et~al.}(1999){Leitherer}, {Schaerer}, {Goldader},
  {Gonz{\'a}lez Delgado}, {Robert}, {Kune}, {de Mello}, {Devost}, \&
  {Heckman}}]{Leitherer1999}
{Leitherer}, C., {Schaerer}, D., {Goldader}, J.~D., {et~al.} 1999, \apjs, 123,
  3

\bibitem[{{Livermore} {et~al.}(2017){Livermore}, {Finkelstein}, \&
  {Lotz}}]{Livermore2017}
{Livermore}, R.~C., {Finkelstein}, S.~L., \& {Lotz}, J.~M. 2017, \apj, 835, 113

\bibitem[{{Lu} {et~al.}(2025){Lu}, {Frenk}, {Bose}, {Lacey}, {Cole}, {Baugh},
  \& {Helly}}]{Lu2025}
{Lu}, S., {Frenk}, C.~S., {Bose}, S., {et~al.} 2025, \mnras, 536, 1018

\bibitem[{{Malhotra} \& {Rhoads}(2004)}]{Malhotra2004}
{Malhotra}, S. \& {Rhoads}, J.~E. 2004, \apjl, 617, L5

\bibitem[{{Martins} \& {Palacios}(2022)}]{Martins2022}
{Martins}, F. \& {Palacios}, A. 2022, \aap, 659, A163

\bibitem[{{Martins} {et~al.}(2023){Martins}, {Schaerer}, {Marques-Chaves}, \&
  {Upadhyaya}}]{Martins2023}
{Martins}, F., {Schaerer}, D., {Marques-Chaves}, R., \& {Upadhyaya}, A. 2023,
  \aap, 678, A159

\bibitem[{{Mascia} {et~al.}(2024){Mascia}, {Pentericci}, {Calabr{\`o}},
  {Santini}, {Napolitano}, {Arrabal Haro}, {Castellano}, {Dickinson}, {Ocvirk},
  {Lewis}, {Amor{\'\i}n}, {Bagley}, {Bhatawdekar}, {Cleri}, {Costantin},
  {Dekel}, {Finkelstein}, {Fontana}, {Giavalisco}, {Grogin}, {Hathi},
  {Hirschmann}, {Holwerda}, {Jung}, {Kartaltepe}, {Koekemoer}, {Lucas},
  {Papovich}, {P{\'e}rez-Gonz{\'a}lez}, {Pirzkal}, {Trump}, {Wilkins}, \&
  {Yung}}]{Mascia2024}
{Mascia}, S., {Pentericci}, L., {Calabr{\`o}}, A., {et~al.} 2024, \aap, 685, A3

\bibitem[{{Mason} {et~al.}(2019){Mason}, {Fontana}, {Treu}, {Schmidt}, {Hoag},
  {Abramson}, {Amorin}, {Brada{\v{c}}}, {Guaita}, {Jones}, {Henry}, {Malkan},
  {Pentericci}, {Trenti}, \& {Vanzella}}]{Mason2019}
{Mason}, C.~A., {Fontana}, A., {Treu}, T., {et~al.} 2019, \mnras, 485, 3947

\bibitem[{{Mason} {et~al.}(2023){Mason}, {Trenti}, \& {Treu}}]{Mason2023}
{Mason}, C.~A., {Trenti}, M., \& {Treu}, T. 2023, \mnras, 521, 497

\bibitem[{{Mason} {et~al.}(2018){Mason}, {Treu}, {Dijkstra}, {Mesinger},
  {Trenti}, {Pentericci}, {de Barros}, \& {Vanzella}}]{Mason2018a}
{Mason}, C.~A., {Treu}, T., {Dijkstra}, M., {et~al.} 2018, \apj, 856, 2

\bibitem[{{Mauerhofer} \& {Dayal}(2023)}]{Mauerhofer2023}
{Mauerhofer}, V. \& {Dayal}, P. 2023, \mnras, 526, 2196

\bibitem[{{McGreer} {et~al.}(2015){McGreer}, {Mesinger}, \&
  {D'Odorico}}]{McGreer2015}
{McGreer}, I.~D., {Mesinger}, A., \& {D'Odorico}, V. 2015, \mnras, 447, 499

\bibitem[{{McLeod} {et~al.}(2024){McLeod}, {Donnan}, {McLure}, {Dunlop},
  {Magee}, {Begley}, {Carnall}, {Cullen}, {Ellis}, {Hamadouche}, \&
  {Stanton}}]{McLeod2024}
{McLeod}, D.~J., {Donnan}, C.~T., {McLure}, R.~J., {et~al.} 2024, \mnras, 527,
  5004

\bibitem[{{McLeod} {et~al.}(2016){McLeod}, {McLure}, \& {Dunlop}}]{McLeod2016}
{McLeod}, D.~J., {McLure}, R.~J., \& {Dunlop}, J.~S. 2016, \mnras, 459, 3812

\bibitem[{{McLeod} {et~al.}(2015){McLeod}, {McLure}, {Dunlop}, {Robertson},
  {Ellis}, \& {Targett}}]{McLeod2015}
{McLeod}, D.~J., {McLure}, R.~J., {Dunlop}, J.~S., {et~al.} 2015, \mnras, 450,
  3032

\bibitem[{{McLure} {et~al.}(2013){McLure}, {Dunlop}, {Bowler}, {Curtis-Lake},
  {Schenker}, {Ellis}, {Robertson}, {Koekemoer}, {Rogers}, {Ono}, {Ouchi},
  {Charlot}, {Wild}, {Stark}, {Furlanetto}, {Cirasuolo}, \&
  {Targett}}]{McLure2013}
{McLure}, R.~J., {Dunlop}, J.~S., {Bowler}, R.~A.~A., {et~al.} 2013, \mnras,
  432, 2696

\bibitem[{{Menon} {et~al.}(2024){Menon}, {Lancaster}, {Burkhart}, {Somerville},
  {Dekel}, \& {Krumholz}}]{Menon2024}
{Menon}, S.~H., {Lancaster}, L., {Burkhart}, B., {et~al.} 2024, \apjl, 967, L28

\bibitem[{{Meurer} {et~al.}(2009){Meurer}, {Wong}, {Kim}, {Hanish}, {Heckman},
  {Werk}, {Bland-Hawthorn}, {Dopita}, {Zwaan}, {Koribalski}, {Seibert},
  {Thilker}, {Ferguson}, {Webster}, {Putman}, {Knezek}, {Doyle}, {Drinkwater},
  {Hoopes}, {Kilborn}, {Meyer}, {Ryan-Weber}, {Smith}, \&
  {Staveley-Smith}}]{Meurer2009}
{Meurer}, G.~R., {Wong}, O.~I., {Kim}, J.~H., {et~al.} 2009, \apj, 695, 765

\bibitem[{{Me{\v{s}}tri{\'c}} {et~al.}(2023){Me{\v{s}}tri{\'c}}, {Vanzella},
  {Upadhyaya}, {Martins}, {Marques-Chaves}, {Schaerer}, {Guibert}, {Zanella},
  {Grillo}, {Rosati}, {Calura}, {Caminha}, {Bolamperti}, {Meneghetti},
  {Bergamini}, {Mercurio}, {Nonino}, \& {Pascale}}]{Mestric2023}
{Me{\v{s}}tri{\'c}}, U., {Vanzella}, E., {Upadhyaya}, A., {et~al.} 2023, \aap,
  673, A50

\bibitem[{{Mirocha} \& {Furlanetto}(2023)}]{Mirocha2023}
{Mirocha}, J. \& {Furlanetto}, S.~R. 2023, \mnras, 519, 843

\bibitem[{{Mu{\~n}oz} {et~al.}(2023){Mu{\~n}oz}, {Mirocha}, {Furlanetto}, \&
  {Sabti}}]{Munoz2023}
{Mu{\~n}oz}, J.~B., {Mirocha}, J., {Furlanetto}, S., \& {Sabti}, N. 2023,
  \mnras, 526, L47

\bibitem[{{Nagele} \& {Umeda}(2023)}]{Nagele2023}
{Nagele}, C. \& {Umeda}, H. 2023, \apjl, 949, L16

\bibitem[{{Nakane} {et~al.}(2024){Nakane}, {Ouchi}, {Nakajima}, {Harikane},
  {Tominaga}, {Takahashi}, {Kashino}, {Yanagisawa}, {Watanabe}, {Nomoto},
  {Isobe}, {Nishigaki}, {Ishigaki}, {Ono}, \& {Takeda}}]{Nakane2024}
{Nakane}, M., {Ouchi}, M., {Nakajima}, K., {et~al.} 2024, \apj, 976, 122

\bibitem[{{Nandal} {et~al.}(2024){Nandal}, {Sibony}, \&
  {Tsiatsiou}}]{Nandal2024}
{Nandal}, D., {Sibony}, Y., \& {Tsiatsiou}, S. 2024, \aap, 688, A142

\bibitem[{{Oesch} {et~al.}(2018){Oesch}, {Bouwens}, {Illingworth}, {Labb{\'e}},
  \& {Stefanon}}]{Oesch2018}
{Oesch}, P.~A., {Bouwens}, R.~J., {Illingworth}, G.~D., {Labb{\'e}}, I., \&
  {Stefanon}, M. 2018, \apj, 855, 105

\bibitem[{{Oesch} {et~al.}(2016){Oesch}, {Brammer}, {van Dokkum},
  {Illingworth}, {Bouwens}, {Labb{\'e}}, {Franx}, {Momcheva}, {Ashby}, {Fazio},
  {Gonzalez}, {Holden}, {Magee}, {Skelton}, {Smit}, {Spitler}, {Trenti}, \&
  {Willner}}]{Oesch2016}
{Oesch}, P.~A., {Brammer}, G., {van Dokkum}, P.~G., {et~al.} 2016, \apj, 819,
  129

\bibitem[{{Oke} \& {Gunn}(1983)}]{Oke1983}
{Oke}, J.~B. \& {Gunn}, J.~E. 1983, \apj, 266, 713

\bibitem[{Oliphant(2006)}]{numpy}
Oliphant, T. 2006, {NumPy}: A guide to {NumPy}, USA: Trelgol Publishing,
  [Online; accessed <today>]

\bibitem[{{Ono} {et~al.}(2012){Ono}, {Ouchi}, {Mobasher}, {Dickinson},
  {Penner}, {Shimasaku}, {Weiner}, {Kartaltepe}, {Nakajima}, {Nayyeri},
  {Stern}, {Kashikawa}, \& {Spinrad}}]{Ono2012}
{Ono}, Y., {Ouchi}, M., {Mobasher}, B., {et~al.} 2012, \apj, 744, 83

\bibitem[{{Ota} {et~al.}(2010){Ota}, {Iye}, {Kashikawa}, {Shimasaku}, {Ouchi},
  {Totani}, {Kobayashi}, {Nagashima}, {Harayama}, {Kodaka}, {Morokuma},
  {Furusawa}, {Tajitsu}, \& {Hattori}}]{Ota2010}
{Ota}, K., {Iye}, M., {Kashikawa}, N., {et~al.} 2010, \apj, 722, 803

\bibitem[{{Ouchi} {et~al.}(2010){Ouchi}, {Shimasaku}, {Furusawa}, {Saito},
  {Yoshida}, {Akiyama}, {Ono}, {Yamada}, {Ota}, {Kashikawa}, {Iye}, {Kodama},
  {Okamura}, {Simpson}, \& {Yoshida}}]{Ouchi2010}
{Ouchi}, M., {Shimasaku}, K., {Furusawa}, H., {et~al.} 2010, \apj, 723, 869

\bibitem[{{Pacucci} {et~al.}(2022){Pacucci}, {Dayal}, {Harikane}, {Inoue}, \&
  {Loeb}}]{Pacucci2022}
{Pacucci}, F., {Dayal}, P., {Harikane}, Y., {Inoue}, A.~K., \& {Loeb}, A. 2022,
  \mnras, 514, L6

\bibitem[{{Papadopoulos} {et~al.}(2011){Papadopoulos}, {Thi}, {Miniati}, \&
  {Viti}}]{Papadopoulos2011}
{Papadopoulos}, P.~P., {Thi}, W.-F., {Miniati}, F., \& {Viti}, S. 2011, \mnras,
  414, 1705

\bibitem[{{Patton} {et~al.}(2020){Patton}, {Wilson}, {Metrow}, {Ellison},
  {Torrey}, {Brown}, {Hani}, {McAlpine}, {Moreno}, \& {Woo}}]{Patton2020}
{Patton}, D.~R., {Wilson}, K.~D., {Metrow}, C.~J., {et~al.} 2020, \mnras, 494,
  4969

\bibitem[{{Pentericci} {et~al.}(2011){Pentericci}, {Fontana}, {Vanzella},
  {Castellano}, {Grazian}, {Dijkstra}, {Boutsia}, {Cristiani}, {Dickinson},
  {Giallongo}, {Giavalisco}, {Maiolino}, {Moorwood}, {Paris}, \&
  {Santini}}]{Pentericci2011}
{Pentericci}, L., {Fontana}, A., {Vanzella}, E., {et~al.} 2011, \apj, 743, 132

\bibitem[{{Pentericci} {et~al.}(2014){Pentericci}, {Vanzella}, {Fontana},
  {Castellano}, {Treu}, {Mesinger}, {Dijkstra}, {Grazian}, {Brada{\v{c}}},
  {Conselice}, {Cristiani}, {Dunlop}, {Galametz}, {Giavalisco}, {Giallongo},
  {Koekemoer}, {McLure}, {Maiolino}, {Paris}, \& {Santini}}]{Pentericci2014}
{Pentericci}, L., {Vanzella}, E., {Fontana}, A., {et~al.} 2014, \apj, 793, 113

\bibitem[{{P{\'e}rez-Gonz{\'a}lez} {et~al.}(2023){P{\'e}rez-Gonz{\'a}lez},
  {Costantin}, {Langeroodi}, {Rinaldi}, {Annunziatella}, {Ilbert}, {Colina},
  {N{\o}rgaard-Nielsen}, {Greve}, {{\"O}stlin}, {Wright}, {Alonso-Herrero},
  {{\'A}lvarez-M{\'a}rquez}, {Caputi}, {Eckart}, {Le F{\`e}vre}, {Labiano},
  {Garc{\'\i}a-Mar{\'\i}n}, {Hjorth}, {Kendrew}, {Pye}, {Tikkanen}, {van der
  Werf}, {Walter}, {Ward}, {Bik}, {Boogaard}, {Bosman}, {G{\'o}mez}, {Gillman},
  {Iani}, {Jermann}, {Melinder}, {Meyer}, {Moutard}, {van Dishoek}, {Henning},
  {Lagage}, {Guedel}, {Peissker}, {Ray}, {Vandenbussche},
  {Garc{\'\i}a-Argum{\'a}nez}, \& {Mar{\'\i}a M{\'e}rida}}]{PerezGonzalez2023}
{P{\'e}rez-Gonz{\'a}lez}, P.~G., {Costantin}, L., {Langeroodi}, D., {et~al.}
  2023, \apjl, 951, L1

\bibitem[{{Planck Collaboration} {et~al.}(2020){Planck Collaboration},
  {Aghanim}, {Akrami}, {Ashdown}, {Aumont}, {Baccigalupi}, {Ballardini},
  {Banday}, {Barreiro}, {Bartolo}, {Basak}, {Battye}, {Benabed}, {Bernard},
  {Bersanelli}, {Bielewicz}, {Bock}, {Bond}, {Borrill}, {Bouchet}, {Boulanger},
  {Bucher}, {Burigana}, {Butler}, {Calabrese}, {Cardoso}, {Carron},
  {Challinor}, {Chiang}, {Chluba}, {Colombo}, {Combet}, {Contreras}, {Crill},
  {Cuttaia}, {de Bernardis}, {de Zotti}, {Delabrouille}, {Delouis}, {Di
  Valentino}, {Diego}, {Dor{\'e}}, {Douspis}, {Ducout}, {Dupac}, {Dusini},
  {Efstathiou}, {Elsner}, {En{\ss}lin}, {Eriksen}, {Fantaye}, {Farhang},
  {Fergusson}, {Fernandez-Cobos}, {Finelli}, {Forastieri}, {Frailis},
  {Fraisse}, {Franceschi}, {Frolov}, {Galeotta}, {Galli}, {Ganga},
  {G{\'e}nova-Santos}, {Gerbino}, {Ghosh}, {Gonz{\'a}lez-Nuevo}, {G{\'o}rski},
  {Gratton}, {Gruppuso}, {Gudmundsson}, {Hamann}, {Handley}, {Hansen},
  {Herranz}, {Hildebrandt}, {Hivon}, {Huang}, {Jaffe}, {Jones}, {Karakci},
  {Keih{\"a}nen}, {Keskitalo}, {Kiiveri}, {Kim}, {Kisner}, {Knox},
  {Krachmalnicoff}, {Kunz}, {Kurki-Suonio}, {Lagache}, {Lamarre}, {Lasenby},
  {Lattanzi}, {Lawrence}, {Le Jeune}, {Lemos}, {Lesgourgues}, {Levrier},
  {Lewis}, {Liguori}, {Lilje}, {Lilley}, {Lindholm}, {L{\'o}pez-Caniego},
  {Lubin}, {Ma}, {Mac{\'\i}as-P{\'e}rez}, {Maggio}, {Maino}, {Mandolesi},
  {Mangilli}, {Marcos-Caballero}, {Maris}, {Martin}, {Martinelli},
  {Mart{\'\i}nez-Gonz{\'a}lez}, {Matarrese}, {Mauri}, {McEwen}, {Meinhold},
  {Melchiorri}, {Mennella}, {Migliaccio}, {Millea}, {Mitra},
  {Miville-Desch{\^e}nes}, {Molinari}, {Montier}, {Morgante}, {Moss}, {Natoli},
  {N{\o}rgaard-Nielsen}, {Pagano}, {Paoletti}, {Partridge}, {Patanchon},
  {Peiris}, {Perrotta}, {Pettorino}, {Piacentini}, {Polastri}, {Polenta},
  {Puget}, {Rachen}, {Reinecke}, {Remazeilles}, {Renzi}, {Rocha}, {Rosset},
  {Roudier}, {Rubi{\~n}o-Mart{\'\i}n}, {Ruiz-Granados}, {Salvati}, {Sandri},
  {Savelainen}, {Scott}, {Shellard}, {Sirignano}, {Sirri}, {Spencer},
  {Sunyaev}, {Suur-Uski}, {Tauber}, {Tavagnacco}, {Tenti}, {Toffolatti},
  {Tomasi}, {Trombetti}, {Valenziano}, {Valiviita}, {Van Tent}, {Vibert},
  {Vielva}, {Villa}, {Vittorio}, {Wandelt}, {Wehus}, {White}, {White},
  {Zacchei}, \& {Zonca}}]{planck2018}
{Planck Collaboration}, {Aghanim}, N., {Akrami}, Y., {et~al.} 2020, \aap, 641,
  A6

\bibitem[{{Planck Collaboration} {et~al.}(2016){Planck Collaboration},
  {Aghanim}, {Ashdown}, {Aumont}, {Baccigalupi}, {Ballardini}, {Banday},
  {Barreiro}, {Bartolo}, {Basak}, {Battye}, {Benabed}, {Bernard}, {Bersanelli},
  {Bielewicz}, {Bock}, {Bonaldi}, {Bonavera}, {Bond}, {Borrill}, {Bouchet},
  {Boulanger}, {Bucher}, {Burigana}, {Butler}, {Calabrese}, {Cardoso},
  {Carron}, {Challinor}, {Chiang}, {Colombo}, {Combet}, {Comis}, {Coulais},
  {Crill}, {Curto}, {Cuttaia}, {Davis}, {de Bernardis}, {de Rosa}, {de Zotti},
  {Delabrouille}, {Delouis}, {Di Valentino}, {Dickinson}, {Diego}, {Dor{\'e}},
  {Douspis}, {Ducout}, {Dupac}, {Efstathiou}, {Elsner}, {En{\ss}lin},
  {Eriksen}, {Falgarone}, {Fantaye}, {Finelli}, {Forastieri}, {Frailis},
  {Fraisse}, {Franceschi}, {Frolov}, {Galeotta}, {Galli}, {Ganga},
  {G{\'e}nova-Santos}, {Gerbino}, {Ghosh}, {Gonz{\'a}lez-Nuevo}, {G{\'o}rski},
  {Gratton}, {Gruppuso}, {Gudmundsson}, {Hansen}, {Helou},
  {Henrot-Versill{\'e}}, {Herranz}, {Hivon}, {Huang}, {Ili{\'c}}, {Jaffe},
  {Jones}, {Keih{\"a}nen}, {Keskitalo}, {Kisner}, {Knox}, {Krachmalnicoff},
  {Kunz}, {Kurki-Suonio}, {Lagache}, {Lamarre}, {Langer}, {Lasenby},
  {Lattanzi}, {Lawrence}, {Le Jeune}, {Leahy}, {Levrier}, {Liguori}, {Lilje},
  {L{\'o}pez-Caniego}, {Ma}, {Mac{\'\i}as-P{\'e}rez}, {Maggio}, {Mangilli},
  {Maris}, {Martin}, {Mart{\'\i}nez-Gonz{\'a}lez}, {Matarrese}, {Mauri},
  {McEwen}, {Meinhold}, {Melchiorri}, {Mennella}, {Migliaccio},
  {Miville-Desch{\^e}nes}, {Molinari}, {Moneti}, {Montier}, {Morgante}, {Moss},
  {Mottet}, {Naselsky}, {Natoli}, {Oxborrow}, {Pagano}, {Paoletti},
  {Partridge}, {Patanchon}, {Patrizii}, {Perdereau}, {Perotto}, {Pettorino},
  {Piacentini}, {Plaszczynski}, {Polastri}, {Polenta}, {Puget}, {Rachen},
  {Racine}, {Reinecke}, {Remazeilles}, {Renzi}, {Rocha}, {Rossetti}, {Roudier},
  {Rubi{\~n}o-Mart{\'\i}n}, {Ruiz-Granados}, {Salvati}, {Sandri}, {Savelainen},
  {Scott}, {Sirri}, {Sunyaev}, {Suur-Uski}, {Tauber}, {Tenti}, {Toffolatti},
  {Tomasi}, {Tristram}, {Trombetti}, {Valiviita}, {Van Tent}, {Vibert},
  {Vielva}, {Villa}, {Vittorio}, {Wandelt}, {Watson}, {Wehus}, {White},
  {Zacchei}, \& {Zonca}}]{planck2016}
{Planck Collaboration}, {Aghanim}, N., {Ashdown}, M., {et~al.} 2016, \aap, 596,
  A107

\bibitem[{{Prieto-Lyon} {et~al.}(2023){Prieto-Lyon}, {Strait}, {Mason},
  {Brammer}, {Caminha}, {Mercurio}, {Acebron}, {Bergamini}, {Grillo}, {Rosati},
  {Vanzella}, {Castellano}, {Merlin}, {Paris}, {Boyett}, {Calabr{\`o}},
  {Morishita}, {Mascia}, {Pentericci}, {Roberts-Borsani}, {Roy}, {Treu}, \&
  {Vulcani}}]{PrietoLyon2023}
{Prieto-Lyon}, G., {Strait}, V., {Mason}, C.~A., {et~al.} 2023, \aap, 672, A186

\bibitem[{{Rinaldi} {et~al.}(2024){Rinaldi}, {Caputi}, {Iani}, {Costantin},
  {Gillman}, {Perez Gonzalez}, {{\"O}stlin}, {Colina}, {Greve},
  {N{\o}rgard-Nielsen}, {Wright}, {{\'A}lvarez-M{\'a}rquez}, {Eckart},
  {Garc{\'\i}a-Mar{\'\i}n}, {Hjorth}, {Ilbert}, {Kendrew}, {Labiano}, {Le
  F{\`e}vre}, {Pye}, {Tikkanen}, {Walter}, {van der Werf}, {Ward},
  {Annunziatella}, {Azzollini}, {Bik}, {Boogaard}, {Bosman}, {Crespo
  G{\'o}mez}, {Jermann}, {Langeroodi}, {Melinder}, {Meyer}, {Moutard},
  {Peissker}, {van Dishoeck}, {G{\"u}del}, {Henning}, {Lagage}, {Ray},
  {Vandenbussche}, {Waelkens}, \& {Dayal}}]{Rinaldi2024}
{Rinaldi}, P., {Caputi}, K.~I., {Iani}, E., {et~al.} 2024, \apj, 969, 12

\bibitem[{Salpeter(1955)}]{Salpeter}
Salpeter, E.~E. 1955, ApJ, 121, 161

\bibitem[{{Schaerer} {et~al.}(2022){Schaerer}, {Marques-Chaves}, {Barrufet},
  {Oesch}, {Izotov}, {Naidu}, {Guseva}, \& {Brammer}}]{Schaerer2022}
{Schaerer}, D., {Marques-Chaves}, R., {Barrufet}, L., {et~al.} 2022, \aap, 665,
  L4

\bibitem[{{Schenker} {et~al.}(2013){Schenker}, {Robertson}, {Ellis}, {Ono},
  {McLure}, {Dunlop}, {Koekemoer}, {Bowler}, {Ouchi}, {Curtis-Lake}, {Rogers},
  {Schneider}, {Charlot}, {Stark}, {Furlanetto}, \& {Cirasuolo}}]{Schenker2013}
{Schenker}, M.~A., {Robertson}, B.~E., {Ellis}, R.~S., {et~al.} 2013, \apj,
  768, 196

\bibitem[{{Schenker} {et~al.}(2012){Schenker}, {Stark}, {Ellis}, {Robertson},
  {Dunlop}, {McLure}, {Kneib}, \& {Richard}}]{Schenker2012}
{Schenker}, M.~A., {Stark}, D.~P., {Ellis}, R.~S., {et~al.} 2012, \apj, 744,
  179

\bibitem[{{Senchyna} {et~al.}(2024){Senchyna}, {Plat}, {Stark}, {Rudie},
  {Berg}, {Charlot}, {James}, \& {Mingozzi}}]{Senchyna2024}
{Senchyna}, P., {Plat}, A., {Stark}, D.~P., {et~al.} 2024, \apj, 966, 92

\bibitem[{{Shen} {et~al.}(2023){Shen}, {Vogelsberger}, {Boylan-Kolchin},
  {Tacchella}, \& {Kannan}}]{Shen2023}
{Shen}, X., {Vogelsberger}, M., {Boylan-Kolchin}, M., {Tacchella}, S., \&
  {Kannan}, R. 2023, \mnras, 525, 3254

\bibitem[{{Simmonds} {et~al.}(2024{\natexlab{a}}){Simmonds}, {Tacchella},
  {Hainline}, {Johnson}, {McClymont}, {Robertson}, {Saxena}, {Sun}, {Witten},
  {Baker}, {Bhatawdekar}, {Boyett}, {Bunker}, {Charlot}, {Curtis-Lake},
  {Egami}, {Eisenstein}, {Hausen}, {Maiolino}, {Maseda}, {Scholtz}, {Williams},
  {Willott}, \& {Witstok}}]{Simmonds2024a}
{Simmonds}, C., {Tacchella}, S., {Hainline}, K., {et~al.} 2024{\natexlab{a}},
  \mnras, 527, 6139

\bibitem[{{Simmonds} {et~al.}(2024{\natexlab{b}}){Simmonds}, {Tacchella},
  {Hainline}, {Johnson}, {Pusk{\'a}s}, {Robertson}, {Baker}, {Bhatawdekar},
  {Boyett}, {Bunker}, {Cargile}, {Carniani}, {Chevallard}, {Curti},
  {Curtis-Lake}, {Ji}, {Jones}, {Kumari}, {Laseter}, {Maiolino}, {Maseda},
  {Rinaldi}, {Stoffers}, {{\"U}bler}, {Villanueva}, {Williams}, {Willott},
  {Witstok}, \& {Zhu}}]{Simmonds2024c}
{Simmonds}, C., {Tacchella}, S., {Hainline}, K., {et~al.} 2024{\natexlab{b}},
  \mnras, 535, 2998

\bibitem[{{Song} {et~al.}(2016){Song}, {Finkelstein}, {Ashby}, {Grazian}, {Lu},
  {Papovich}, {Salmon}, {Somerville}, {Dickinson}, {Duncan}, {Faber}, {Fazio},
  {Ferguson}, {Fontana}, {Guo}, {Hathi}, {Lee}, {Merlin}, \&
  {Willner}}]{song_evolution_2016}
{Song}, M., {Finkelstein}, S.~L., {Ashby}, M. L.~N., {et~al.} 2016, \apj, 825,
  5

\bibitem[{{Sun} {et~al.}(2023){Sun}, {Faucher-Gigu{\`e}re}, {Hayward}, {Shen},
  {Wetzel}, \& {Cochrane}}]{Sun2023}
{Sun}, G., {Faucher-Gigu{\`e}re}, C.-A., {Hayward}, C.~C., {et~al.} 2023,
  \apjl, 955, L35

\bibitem[{{Topping} {et~al.}(2024){Topping}, {Stark}, {Senchyna}, {Chen},
  {Zitrin}, {Endsley}, {Charlot}, {Furtak}, {Maseda}, {Plat}, {Smit},
  {Mainali}, {Chevallard}, {Molyneux}, \& {Rigby}}]{Topping2024}
{Topping}, M.~W., {Stark}, D.~P., {Senchyna}, P., {et~al.} 2024, arXiv
  e-prints, arXiv:2407.19009

\bibitem[{{Totani} {et~al.}(2014){Totani}, {Aoki}, {Hattori}, {Kosugi},
  {Niino}, {Hashimoto}, {Kawai}, {Ohta}, {Sakamoto}, \& {Yamada}}]{Totani2014}
{Totani}, T., {Aoki}, K., {Hattori}, T., {et~al.} 2014, \pasj, 66, 63

\bibitem[{{Totani} {et~al.}(2006){Totani}, {Kawai}, {Kosugi}, {Aoki}, {Yamada},
  {Iye}, {Ohta}, \& {Hattori}}]{Totani2006}
{Totani}, T., {Kawai}, N., {Kosugi}, G., {et~al.} 2006, \pasj, 58, 485

\bibitem[{{Treu} {et~al.}(2012){Treu}, {Trenti}, {Stiavelli}, {Auger}, \&
  {Bradley}}]{Treu2012}
{Treu}, T., {Trenti}, M., {Stiavelli}, M., {Auger}, M.~W., \& {Bradley}, L.~D.
  2012, \apj, 747, 27

\bibitem[{{Trinca} {et~al.}(2024){Trinca}, {Schneider}, {Valiante}, {Graziani},
  {Ferrotti}, {Omukai}, \& {Chon}}]{Trinca2024}
{Trinca}, A., {Schneider}, R., {Valiante}, R., {et~al.} 2024, \mnras, 529, 3563

\bibitem[{{Ucci} {et~al.}(2023){Ucci}, {Dayal}, {Hutter}, {Kobayashi},
  {Gottl{\"o}ber}, {Yepes}, {Hunt}, {Legrand}, \& {Tortora}}]{Ucci2023}
{Ucci}, G., {Dayal}, P., {Hutter}, A., {et~al.} 2023, \mnras, 518, 3557

\bibitem[{{Upadhyaya} {et~al.}(2024){Upadhyaya}, {Marques-Chaves}, {Schaerer},
  {Martins}, {P{\'e}rez-Fournon}, {Palacios}, \& {Stanway}}]{Upadhyaya2024}
{Upadhyaya}, A., {Marques-Chaves}, R., {Schaerer}, D., {et~al.} 2024, \aap,
  686, A185

\bibitem[{{van Dokkum} \& {Conroy}(2024)}]{vanDokkumConroy2024}
{van Dokkum}, P. \& {Conroy}, C. 2024, \apjl, 973, L32

\bibitem[{{Vanzella} {et~al.}(2023){Vanzella}, {Loiacono}, {Bergamini},
  {Me{\v{s}}tri{\'c}}, {Castellano}, {Rosati}, {Meneghetti}, {Grillo},
  {Calura}, {Mignoli}, {Brada{\v{c}}}, {Adamo}, {Rihtar{\v{s}}i{\v{c}}},
  {Dickinson}, {Gronke}, {Zanella}, {Annibali}, {Willott}, {Messa}, {Sani},
  {Acebron}, {Bolamperti}, {Comastri}, {Gilli}, {Caputi}, {Ricotti},
  {Gruppioni}, {Ravindranath}, {Mercurio}, {Strait}, {Martis}, {Pascale},
  {Caminha}, {Annunziatella}, \& {Nonino}}]{Vanzella2023}
{Vanzella}, E., {Loiacono}, F., {Bergamini}, P., {et~al.} 2023, \aap, 678, A173

\bibitem[{{Vink}(2023)}]{Vink2023}
{Vink}, J.~S. 2023, \aap, 679, L9

\bibitem[{{Wang} {et~al.}(2023){Wang}, {Fujimoto}, {Labb{\'e}}, {Furtak},
  {Miller}, {Setton}, {Zitrin}, {Atek}, {Bezanson}, {Brammer}, {Leja}, {Oesch},
  {Price}, {Chemerynska}, {Cutler}, {Dayal}, {van Dokkum}, {Goulding},
  {Greene}, {Fudamoto}, {Khullar}, {Kokorev}, {Marchesini}, {Pan}, {Weaver},
  {Whitaker}, \& {Williams}}]{Wang2023}
{Wang}, B., {Fujimoto}, S., {Labb{\'e}}, I., {et~al.} 2023, \apjl, 957, L34

\bibitem[{{Watanabe} {et~al.}(2024{\natexlab{a}}){Watanabe}, {Ouchi}, {Isobe},
  \& {Tominaga}}]{Watanabe2024a}
{Watanabe}, K., {Ouchi}, M., {Isobe}, Y., \& {Tominaga}, N. 2024{\natexlab{a}},
  in EAS2024, 1243

\bibitem[{{Watanabe} {et~al.}(2024{\natexlab{b}}){Watanabe}, {Ouchi},
  {Nakajima}, {Isobe}, {Tominaga}, {Suzuki}, {Ishigaki}, {Nomoto}, {Takahashi},
  {Harikane}, {Hatano}, {Kusakabe}, {Moriya}, {Nishigaki}, {Ono}, {Onodera}, \&
  {Sugahara}}]{Watanabe2024}
{Watanabe}, K., {Ouchi}, M., {Nakajima}, K., {et~al.} 2024{\natexlab{b}}, \apj,
  962, 50

\bibitem[{{Watts} {et~al.}(2018){Watts}, {Meurer}, {Lagos}, {Bruzzese},
  {Kroupa}, \& {Jerabkova}}]{Watts2018}
{Watts}, A.~B., {Meurer}, G.~R., {Lagos}, C. D.~P., {et~al.} 2018, \mnras, 477,
  5554

\bibitem[{{Weidner} {et~al.}(2013){Weidner}, {Kroupa}, {Pflamm-Altenburg}, \&
  {Vazdekis}}]{Weidner2013}
{Weidner}, C., {Kroupa}, P., {Pflamm-Altenburg}, J., \& {Vazdekis}, A. 2013,
  \mnras, 436, 3309

\bibitem[{{Whitler} {et~al.}(2024){Whitler}, {Stark}, {Endsley}, {Chen},
  {Mason}, {Topping}, \& {Charlot}}]{Whitler2024}
{Whitler}, L., {Stark}, D.~P., {Endsley}, R., {et~al.} 2024, \mnras, 529, 855

\bibitem[{{Witstok} {et~al.}(2024){Witstok}, {Jakobsen}, {Maiolino}, {Helton},
  {Johnson}, {Robertson}, {Tacchella}, {Cameron}, {Smit}, {Bunker}, {Saxena},
  {Sun}, {Arribas}, {Baker}, {Bhatawdekar}, {Boyett}, {Cargile}, {Carniani},
  {Charlot}, {Chevallard}, {Curti}, {Curtis-Lake}, {D'Eugenio}, {Eisenstein},
  {Hainline}, {Jones}, {Kumari}, {Maseda}, {P{\'e}rez-Gonz{\'a}lez}, {Rinaldi},
  {Scholtz}, {{\"U}bler}, {Williams}, {Willmer}, {Willott}, \&
  {Zhu}}]{Witstok2024}
{Witstok}, J., {Jakobsen}, P., {Maiolino}, R., {et~al.} 2024, arXiv e-prints,
  arXiv:2408.16608

\bibitem[{{Wofford} {et~al.}(2023){Wofford}, {Sixtos}, {Charlot}, {Bruzual},
  {Cullen}, {Stanton}, {Hern{\'a}ndez}, {Smith}, \& {Hayes}}]{Wofford2023}
{Wofford}, A., {Sixtos}, A., {Charlot}, S., {et~al.} 2023, \mnras, 523, 3949

\bibitem[{{Yan} {et~al.}(2017){Yan}, {Jerabkova}, \& {Kroupa}}]{Yan2017}
{Yan}, Z., {Jerabkova}, T., \& {Kroupa}, P. 2017, \aap, 607, A126

\bibitem[{{Yoshida} {et~al.}(2006){Yoshida}, {Omukai}, {Hernquist}, \&
  {Abel}}]{Yoshida2006}
{Yoshida}, N., {Omukai}, K., {Hernquist}, L., \& {Abel}, T. 2006, \apj, 652, 6

\bibitem[{{Yung} {et~al.}(2024){Yung}, {Somerville}, {Finkelstein}, {Wilkins},
  \& {Gardner}}]{Yung2023}
{Yung}, L.~Y.~A., {Somerville}, R.~S., {Finkelstein}, S.~L., {Wilkins}, S.~M.,
  \& {Gardner}, J.~P. 2024, \mnras, 527, 5929

\bibitem[{{Zhang} {et~al.}(2018){Zhang}, {Romano}, {Ivison}, {Papadopoulos}, \&
  {Matteucci}}]{Zhang2018}
{Zhang}, Z.-Y., {Romano}, D., {Ivison}, R.~J., {Papadopoulos}, P.~P., \&
  {Matteucci}, F. 2018, \nat, 558, 260

\bibitem[{{Ziparo} {et~al.}(2023){Ziparo}, {Ferrara}, {Sommovigo}, \&
  {Kohandel}}]{Ziparo2023}
{Ziparo}, F., {Ferrara}, A., {Sommovigo}, L., \& {Kohandel}, M. 2023, \mnras,
  520, 2445

\end{thebibliography}

\begin{appendix}

\section{Time evolution of emissivities}
\label{app_fits}

For an IMF described by Eqn.\ref{eq_evolvingIMF}, we derive fitting functions to model the time evolution of ionising and ultraviolet emissivities for a $1\msun$ starburst, considering all available combinations of $f_\mathrm{massive}$ and $Z$ values ($f_\mathrm{massive} \in [0,1]$, $Z\in[0.001, 0.008]$). For metallicity values lower or larger than the covered interval the values for the lower or upper limit are assumed, respectively.

\subsection{Ionising emissivity}

We approximate the time evolution of the number of photons above an energy of $13.6$~eV per second for a starburst of $1~\msun$ with the following fitting function
\begin{eqnarray}
    \frac{N_\mathrm{ion}}{\mathrm{s}^{-1}} = 
    \begin{cases}
        N_0 &, \ t\leq t_\mathrm{break}\\
        N_0\ \left(t/t_\mathrm{break}\right)^{-\gamma} &, \ t>t_\mathrm{break},
    \end{cases}
\end{eqnarray}
where the normalisation $N_0$, the break $t_\mathrm{break}$, and slope $\gamma$ are given by
{\small
\begin{eqnarray}
    \lg \left(\frac{t_\mathrm{break}}{\mathrm{Myr}}\right) =
    \begin{cases}
        \max\left[0.49 ~\left(1 - 23.37 Z\right) + 2.64~\left(f_\mathrm{massive} - 0.209\right)^2, 0.477\right], \\
        \hspace{5.6cm} f_\mathrm{massive}<0.209 \\
        \max\left[0.47 ~(1 -17.15 Z) + 0.06~(1 - e^{-7.5~(f_\mathrm{massive} - 0.212)}, 0.477\right], \\
        \hspace{5.6cm} f_\mathrm{massive} \geq 0.209
    \end{cases}
\end{eqnarray}
\begin{eqnarray}
    \gamma =
    \begin{cases}
        -5.07~(1+2.16 Z) - 0.49~(1+8.96 Z), \hspace{0.8cm} f_\mathrm{massive} \geq 0.7 - 0.5Z\\
        -5.07~(1+2.16 Z) + 0.49~(1+8.96 Z) \\
        \ \ \ \ \times \cos\left[-5.15 f_\mathrm{massive} - 6.16 (1 + 2048 Z)\right], \ \ f_\mathrm{massive} < 0.7 - 0.5Z
    \end{cases}
\end{eqnarray}
\begin{eqnarray}
    \lg N_0 &=& 3.4\times10^{-9} (1+0.62Z) + 53.66~ (1-0.32Z) \nonumber\\
    && \times \left[f_\mathrm{massive} + 0.051~(1+227.6Z) \right]^{0.00766~(1+50.01Z)}.
\end{eqnarray}
}

\subsection{Ultraviolet luminosity}

We approximate the time evolution of the UV luminosity for a starburst of $1~\msun$ with the following fitting function
\begin{eqnarray}
    \lg \left(\frac{L_\mathrm{UV}}{\mathrm{~erg~s}^{-1}\mathrm{\AA}^{-1}}\right) =
    \begin{cases}
        \lg N_0, \hspace{3.3cm} t \leq 1.26~\mathrm{Myr}\\
        \lg N_0 + 26.08~ \lg N_1 \lg\left(\frac{t}{1.26~\mathrm{Myr}}\right), \\ 
        \hspace{2.6cm} 1.26~\mathrm{Myr} < t \leq 3.5~\mathrm{Myr}\\
        \lg N_0 +  \lg N_2 + \gamma ~ \lg\left(\frac{t}{3.5~\mathrm{Myr}}\right), \ t > 3.5~\mathrm{Myr}
    \end{cases}
\end{eqnarray}
where $\gamma$, $N_0$, $N_1$ and $N_2$ are given by
{\small
\begin{eqnarray}
    \gamma =
    \begin{cases}
        -1.920 + 0.123 e^{-2\times10^3 (Z-0.001)} - 0.564, \hspace{1.6cm} f_\mathrm{massive} \leq 0.7 \\
        -1.920 + 0.123 e^{-2\times10^3 (Z-0.001)} + 0.564 \cos(-4.155 f_\mathrm{massive} - 0.385),\\
        \hspace{6.5cm} f_\mathrm{massive} > 0.7 \\
    \end{cases} 
\end{eqnarray}
\begin{eqnarray}
    \lg N_0 = 36.0 ~ (1 + 0.1 Z) + 3.82 ~ (1 + 1.1 Z) ~ (f_\mathrm{massive} + 0.13)^{0.146} - 6
\end{eqnarray}
\begin{eqnarray}
    \lg N_2 =
    \begin{cases}
        \max\left[ -0.072~(1 + 98.64 Z) + 4.92~(f_\mathrm{massive} - 0.252)^2, 0 \right],\\
        \hspace{5.8cm} f_\mathrm{massive}<0.252 \\        
        0, \hspace{4.4cm} 0.252 \leq f_\mathrm{massive} \leq 0.452 \\
        \max\left[0.023~(1 + -53691 Z^2) + 0.06~\left(1 - e^{-7.5~(f_\mathrm{massive} - 0.452)}\right), 0 \right],\\
        \hspace{5.8cm} f_\mathrm{massive} > 0.452 
    \end{cases}
\end{eqnarray}
\begin{eqnarray}
    \lg N_1 =
    \begin{cases}
       0.064~\left[1 + 8166(Z-0.0035)^2\right] ~ e^{-3.29~\left[f_\mathrm{massive} - 0.5~(1 + 9.89~(Z-0.006)^2)\right]^2},\\
       \hspace{5.8cm} f_\mathrm{massive}\leq 0.283 \\   
      0.044~\left[1 + 9311(Z-0.0041)^2\right] + 0.009~\left[1 + 24847~(Z-0.002)^2\right] \\
      \ \ \ \ \times ~ e^{-4.5~(1 + 99886~(Z-0.006)^2) ~(f_\mathrm{massive} - 0.283)}, \hspace{1.cm} f_\mathrm{massive} > 0.283.    
    \end{cases}
\end{eqnarray}
}

\end{appendix}

\end{document}